\documentclass[a4paper,twoside,final,12pt,titlepage]{article}%
\usepackage{amsmath}
\usepackage{amsfonts}
\usepackage{amssymb}
\usepackage{graphicx}%
\providecommand{\U}[1]{\protect\rule{.1in}{.1in}}
\newtheorem{theorem}{Theorem}

\newtheorem{proposition}[theorem]{Proposition}

\newenvironment{proof}[1][Proof]{\noindent\textbf{#1.} }{\ \rule{0.5em}{0.5em}}
\newskip\humongous \humongous=0pt plus 1000pt minus 1000pt

\newif\ifdtup

\catcode`\@=11
\@addtoreset{equation}{section}

\def\@normalsize{\@setsize\normalsize{15pt}\xiipt\@xiipt
\abovedisplayskip 14pt plus3pt minus3pt\belowdisplayskip \abovedisplayskip
\abovedisplayshortskip \z@ plus3pt\belowdisplayshortskip 7pt plus3.5pt minus0pt}
\def\small{\@setsize\small{13.6pt}\xipt\@xipt
\abovedisplayskip 13pt plus3pt minus3pt\belowdisplayskip \abovedisplayskip
\abovedisplayshortskip \z@ plus3pt\belowdisplayshortskip 7pt plus3.5pt minus0pt
\def\@listi{\parsep 4.5pt plus 2pt minus 1pt
\itemsep \parsep
\topsep 9pt plus 3pt minus 3pt}}
\relax
\catcode`@=12
\catcode`\@=11
\begin{document}
\begin{titlepage}
\renewcommand{\thefootnote}{\fnsymbol{footnote}}
\bigskip
\bigskip
\bigskip
\bigskip
\bigskip
\begin{center}
{\Large  {\bf Multi-monopoles and Magnetic Bags}
}
\end{center}
\bigskip
\renewcommand{\thefootnote}{\fnsymbol{footnote}}
\bigskip
\begin{center}
{\large   Stefano Bolognesi\footnote{s.bolognesi@nbi.dk} }
\vskip 0.20cm
\end{center}
\begin{center}
{\it      \footnotesize
The Niels Bohr Institute, Blegdamsvej 17, DK-2100 Copenhagen \O, Denmark} \\
\end {center}
\renewcommand{\thefootnote}{\arabic{footnote}}
\setcounter{footnote}{0}
\bigskip
\bigskip
\bigskip
\noindent
\begin{center} {\bf Abstract} \end{center}
By analogy with the multi-vortices, we show that also multi-monopoles
become magnetic bags in the large $n$ limit. This simplification allows us to compute the spectrum and the profile
functions  by requiring the minimization of the energy of the bag. We
consider in detail the case of the  magnetic bag in the limit of vanishing potential and we find that it saturates the Bogomol'nyi
bound and there is an infinite set of different shapes of allowed bags. This is consistent with the existence of a moduli space
of solutions for the BPS multi-monopoles. We discuss the string theory interpretation of our result and also the relation between
the 't Hooft large $n$ limit of certain supersymmetric gauge theories and the large $n$ limit of multi-monopoles.
We then consider multi-monopoles in the cosmological context and provide a mechanism that could lead to their production.
\vfill
\begin{flushleft}
May, 2006
\end{flushleft}
\end{titlepage}

\section{Introduction}

In a recent series of works \cite{wallvortex,Znstrings,proof} we studied the
behavior of the Abrikosov-Nielsen-Olesen (ANO) multi-vortex
\cite{Abrikosov,NielsenOlesen} in the large $n$ limit, where $n$ is the number
of quanta of magnetic flux. In this limit the multi-vortex becomes a wall
vortex, that is a wall compactified on a cylinder and stabilized by the
magnetic flux inside. The wall vortex is essentially a bag, such as the ones
studied in the context of the bag models of hadrons
\cite{Bardeen:1974wr,Chodos:1974je,Friedberg:1977xf}. When we say \emph{bag},
we mean any object that is composed by a wall of tension $T_{\mathrm{W}}$ and
thickness $\Delta_{\mathrm{W}}$, that separates an internal region with energy
density $\varepsilon_{0}$ and an external region with energy density $0$.
There are two kind of forces that comprise the bag: one comes from the tension
of the wall $T_{\mathrm{W}}$ and the other comes from the internal energy
density $\varepsilon_{0}$. If the first dominates we use the name SLAC bag
\cite{Chodos:1974je}, while if the second dominates we use the name MIT bag
\cite{Bardeen:1974wr}. The bag, to be stable, needs also some pressure force
to balance the collapse forces. In the case of the wall vortex this force is
due to the magnetic flux inside the cylinder.

In the present paper we consider the 't Hooft-Polyakov magnetic monopole
\cite{'tHooft:1974qc} and we will see that also in this case the
multi-monopole becomes a bag in the large $n$ limit. This object, which we
from now on will denote \emph{magnetic bag}, is composed by a domain wall of
thickness $1/M_{W}$ (where $M_{W}$ is the mass of the $W^{\pm}$ bosons)
wrapped around a closed surface $\mathcal{S}_{\mathrm{m}}$. The surface
$\mathcal{S}_{\mathrm{m}}$ can be spherical or can also have different shapes.
The shape of the surface encodes in some sense the information about the point
on the moduli space of the multi-monopole. The size of the bag is of order
$n/M_{W}$ so in the large $n$ limit it is much bigger than the thickness of
the wall. In what follows we will consider the simplest model that admits
magnetic monopoles: the $SU(2)$ gauge theory spontaneously broken to $U(1)$ by
an adjoint scalar field. The bag surface $\mathcal{S}_{\mathrm{m}}$ separates
two regions, the internal one that is in the non-abelian $SU(2)$ phase and the
external one that is in the abelian phase. The size of the bag is determined
by the balance of the attractive and repulsive forces. Attractive forces are
due to the internal energy density, the tension of the wall and, in case of
vanishing potential, the long tail of the Higgs field outside the bag. The
repulsive force is instead due to the abelian magnetic field outside the bag.

It is important to say that this paper does not give a rigorous proof that
multi-monopoles become magnetic bags in the large $n$ limit, but only some
evidence for it. For this reason we will sometime refer to our claim as the
\emph{magnetic bag conjecture}. Let us make a parallel with the
multi-vortices. In this case we started in \cite{wallvortex,Znstrings} giving
some evidences for the fact that in the large $n$ limit multi-vortices become
wall vortices. There where two kind of evidence: a qualitative analysis of the
multi-vortex differential equations and the surprising coincidence that the
wall vortex saturates exactly the BPS bound. At this level the idea was only a
conjecture but in \cite{proof} we provided an accurate numerical analysis of
the differential equations that, in our opinion, gave a quite convincing proof
of our assertion. In the present paper we will follow a similar path for the
magnetic bag conjecture. First we will give a motivation from the qualitative
behavior of the Higgs field profile\ at the origin (polynomial $\propto r^{n}%
$) and at infinity (it approaches $1$ exponentially as $\propto e^{-M_{W}r}$).
Then we will find that the magnetic bag saturates exactly the BPS bound. We
will also find that there is an infinite set of different shapes allowed for
the magnetic bag and this matches with the existence of a moduli space of
solutions for the multi-monopole. The last step is to check the conjecture on
some large $n$ solution. One major difference with respect to the multi-vortex
is that there is not a spherically symmetric multi-monopole with charge
greater than one \cite{Weinberg:1976eq}. The maximal possible symmetry is the
axial one that has been studied in great detail in Refs.
\cite{Ward:1981jb,prasad} and there are exact formulas for the profile
functions of the multi-monopoles. Using these solutions we can find a quite
convincing evidence for the magnetic bag conjecture in the case of the axial
symmetric multi-monopole.

In this paper we will also interpret the magnetic bag in the string theory
frame-work. Magnetic monopoles can be realized in string theory as D$1$-branes
suspended between D$3$-branes. A D$1$-brane ending on a single D$3$-brane can
be viewed as a deformation of the world volume of the D$3$-brane. This
deformation, also called BIon, is a spike with profile $\propto1/r$ emerging
from the D$3$-brane and is a solution of the abelian Dirac-Born-Infeld (DBI)
action that describe the low energy degrees of freedom on the D$3$-brane
\cite{Bions}. When considering the D$1$-brane suspended between two
D$3$-branes we have to use the non-abelian generalization of the DBI action
and the D$1$-brane becomes a couple of spikes emerging from the two
D$3$-branes and ending in a common point \cite{hashimoto}. The profile of the
two spikes is the same as that of the Higgs field solution in the BPS 't
Hooft-Polyakov monopole. When we take a large number of D$1$-branes at the
same position, the point where the two spikes are joined together expand out
in a closed surface which is nothing but the boundary of the magnetic bag.
Inside the bag are the two D$3$-branes, one on the top of the other and the
non-abelian $SU(2)$ symmetry is restored. Outside of the bag the two
D$3$-branes are deformed in the same way as a cut abelian BIonic spike. The
\textquotedblleft non-abelianity\textquotedblright survives only in a thin
region of thickness $1/M_{W}$ around the surface of the magnetic bag.

In the last part of the paper we want to explore the possible formation of
multi-monopoles in the cosmological context. Magnetic monopoles are naturally
part of the spectrum of grand unified theories (GUTs). Whenever the grand
unification group is semi-simple, there are 't Hooft-Polyakov monopoles with
typical mass of the GUT scale $10^{15}$ GeV. The only way to produce these
super-heavy defects, is in the cosmological context when the temperature of
the universe was of order of the GUT scale. When the temperature cooled below
the critical GUT temperature, the GUT Higgs boson condensed in causally
disconnected domains. At the intersection of different domains there is some
probability $p$ ( $\sim0.1$) to find a monopole of charge $1$. The probability
to find a multi-monopole at this stage can be neglected. We will find that for
a particular choice of the GUT Higgs potential multi-monopoles can be produced
after the phase transition and, with an extreme choice of parameters, our
mechanism provides also a new possible solution of the cosmological monopole problem.

The paper is organized as follows. In Section \ref{review} we give a brief
review of the 't Hooft-Polyakov monopole. This part contains no original
material, it is just a collection of results that we will use in the rest of
the paper. In Section \ref{bag} we discuss multi-monopoles and magnetic bags
in the large $n$ limit. In Section \ref{modulispace} we study the moduli space
of BPS bags. In Section \ref{moreonthebag} we clarify some aspects of the
magnetic bag conjecture and we find more evidence for it studying the exact
known solution of the axial symmetric multi-monopoles. In Section
\ref{multiandgauge} the relation between the 't Hooft large $n$ limit of
certain supersymmetric gauge theories and the large $n$ limit of
multi-monopoles. In Section \ref{string} we interpret our results in the
context of string theory. We conclude in Section \ref{cosmology} speculating
about the possible production of GUT multi-monopoles in the cosmological context.

\section{Magnetic Monopoles \label{review}}

\begin{figure}
[tbh]
\begin{center}
\includegraphics[
trim=0.000000in 0.000000in -0.002528in 0.000000in,
height=2.3557in,
width=3.0874in
]%
{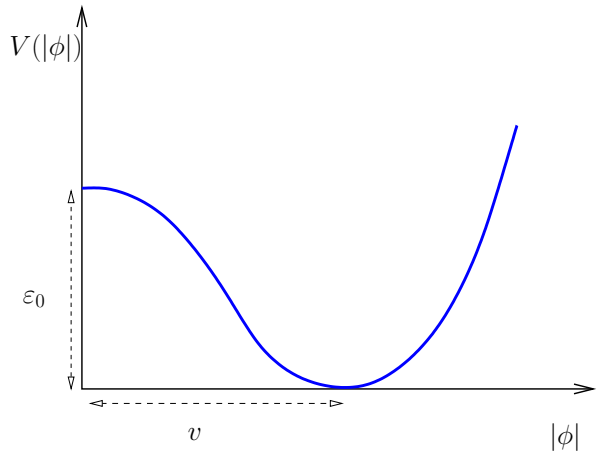}%
\caption{{\protect\footnotesize The Higgs potential}}%
\label{potentialmon}%
\end{center}
\end{figure}

We consider the simplest unified theory that admits magnetic monopoles (see
\cite{Goddard:1977da,Preskill:1984gd} for reviews). It is as $SU(2)$ gauge
theory coupled to a scalar field in the adjoint representation%
\begin{equation}
\mathcal{L}=-\frac{1}{4}F_{\mu\nu}^{a}F^{\mu\nu a}-\frac{1}{2}D_{\mu}\phi
^{a}D^{\mu}\phi^{a}-V(\phi)\ , \label{lagrangian}%
\end{equation}
where the covariant derivative is $D_{\mu}\phi^{a}=\partial_{\mu}\phi
^{a}+ee^{abc}A_{\mu}^{b}\phi^{c}$ and $e$ is the coupling constant. The
potential is%
\begin{equation}
V(\phi)=\frac{1}{8}\lambda(\phi^{a}\phi^{a}-v^{2})^{2}\ . \label{potential}%
\end{equation}
The potential (shown in Figure \ref{potentialmon}) is such that the vacuum
manifold is the sphere $\mathbf{S}^{2}$ of the vectors of fixed norm
$|\phi|=v$.%
The scalar field condensate breaks the gauge group down to $U(1)$. Expanding
around the vacuum we obtain the masses of the perturbative particles,
respectively the $W^{\pm}$ bosons and the Higgs boson:%
\begin{equation}
M_{W}=ev\ ,\qquad M_{H}=\sqrt{\lambda}v\ .
\end{equation}

Due to the non trivial topology of the vacuum manifold, the theory admits a
nonperturbative particle: the 't Hooft-Polyakov magnetic monopole
\cite{'tHooft:1974qc}. It is constructed in the following way. The second
homotopy group of the vacuum manifold is $\pi_{2}(\mathbf{S}^{2})=\mathbb{Z}$
and so we can choose a non-trivial configuration such that the spatial sphere
$\mathbf{S}^{2}$ at infinity is winded once around the vacuum manifold
$\mathbf{S}^{2}$. In order to have a finite energy configuration, we have to
choose the gauge field such that the covariant derivative $D_{\mu}\phi$
vanishes rapidly enough at infinity. The monopole field distribution, in
spherical coordinates, is%
\begin{equation}
\phi^{a}=v\hat{r}^{a}H(M_{W}r)\ ,\qquad A_{i}^{a}=\frac{\epsilon_{iak}\hat
{r}^{k}}{er}(1-K(M_{W}r))\ , \label{ansatz}%
\end{equation}
where the profile functions must satisfy the boundary conditions $H(\infty
)=1$, $K(\infty)=0$ and $H(0)=0$, $K(0)=1$. The profiles $H$ and $K$ are
functions of a dimensionless variable $M_{W}r$, and the radius of the monopole
$R_{\mathrm{m}}$ is of order $\sim1/M_{W}$. The corresponding energy of the
Lagrangian (\ref{lagrangian}) is%
\begin{equation}
E=\int d^{3}r\left[  \frac{1}{2}B_{i}^{a}B_{i}^{a}+\frac{1}{2}D_{i}\phi
^{a}D_{i}\phi^{a}+V(\phi)\right]  \ ,
\end{equation}
where $B_{i}^{a}=\frac{1}{2}\epsilon_{ijk}F_{jk}^{a}$ is the non-abelian
magnetic field. The differential equations to determine the profile functions
are obtained by minimizing the energy functional and the mass of the monopole
is obtained by evaluating the energy at its minimum. The result is%
\begin{equation}
M_{\mathrm{m}}=\frac{4\pi v}{e}f(\lambda/e^{2})\ , \label{mass}%
\end{equation}
where $f$ is a slow varying monotonic function that satisfies $f(0)=1$ and
$f(\infty)=1.787$.

Using the Bogomol'nyi trick \cite{Bogomolny} we can obtain a lower bound on
the mass of the monopole%
\begin{align}
M_{\mathrm{m}}  &  \geq\int d^{3}r\left[  \frac{1}{2}B_{i}^{a}B_{i}^{a}%
+\frac{1}{2}D_{i}\phi^{a}D_{i}\phi^{a}\right] \nonumber\label{bound}\\
&  =\int d^{3}r\frac{1}{2}(B_{i}^{a}\pm D_{i}\phi^{a})^{2}\mp\int d^{3}%
rB_{i}^{a}D_{i}\phi^{a}\nonumber\\
&  \geq\frac{4\pi v}{e}|n|\ .
\end{align}
Where $n$ is the topological degree of the map from $\mathbf{S}^{2}$ at
spatial infinity to the vacuum manifold $\mathbf{S}^{2}$. The limit where the
potential is sent to zero, while keeping fixed the vev $v$, is the so called
BPS limit. In this limit an exact solution for the monopole of charge $1$ was
first obtained by Prasad and Sommerfield in \cite{Prasad:1975kr}. This
solution saturates the Bogomol'nyi bound.

Now we come to multi-monopoles. It was pointed out in \cite{Weinberg:1976eq}
that multi-monopoles with topological charge greater then $1$ cannot be
spherical symmetric. This has to do with the topology of the maps of
$\mathbf{S}^{2}$ onto itself. Call $\phi$ a generic map, it is defined
spherical symmetric if%
\begin{equation}
\phi=g\phi g^{-1}\ ,
\end{equation}
for every choice of the isometry $g\in SO(3)$ of the sphere $\mathbf{S}^{2}$.
The only spherical map that exist is in the topological sector $n=1$ and is
the identity. This is the map of the 't Hooft-Polyakov monopole at infinity
(\ref{ansatz}). The maximal symmetry that we can have for multi-monopoles is
the axial one. An exact solution for the axial symmetric monopole of charge
$2$ has been found by Ward in \cite{Ward:1981jb}. Thereafter Prasad
\cite{prasad} found a method to construct an axial symmetric monopole of
generic charge $n$.

\section{Multi-monopoles are Magnetic Bags \label{bag}}

In this section we are going to consider multi-monopoles in the large $n$
limit. As in the case of multi-vortices, a great simplification occurs when
the number of magnetic flux is large enough. The soliton becomes a bag, which
is a domain wall of negligible thickness that separates an internal region
where $\phi=0$ from an external region where the theory is in the abelian phase. In this limit we can
compute the mass and the profile functions using simple arguments.

\subsection{The magnetic bag}

\begin{figure}
[tbh]
\begin{center}
\includegraphics[
height=2.7864in,
width=3.1877in
]%
{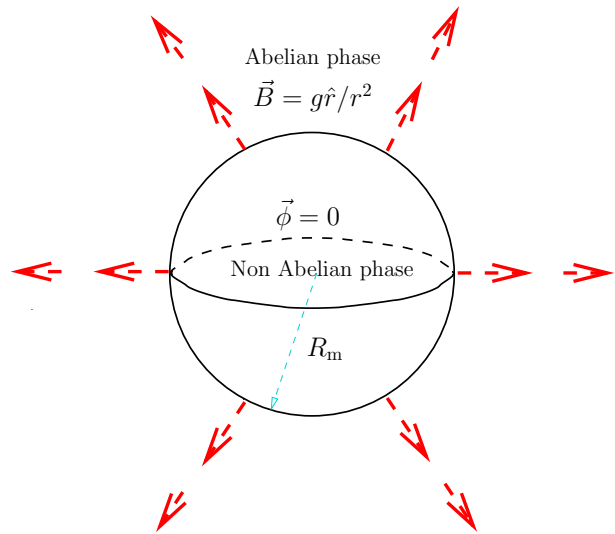}%
\caption{\underline{{\protect\footnotesize The magnetic bag}}%
{\protect\footnotesize . The sphere ${\cal S}^2$  separates two regions:
the internal one is in the non-abelian false vacuum $\phi=0$, the external one
is in the abelian phase. }}%
\label{magneticbag}%
\end{center}
\end{figure}

Our claim is that, for sufficiently large $n$, multi-monopoles become magnetic
bags such as the one shown in Figure \ref{magneticbag}.%
In the internal region $\vec{\phi}=0$. This is an instable stationary point of
the potential in Figure \ref{potentialmon} with energy density $\varepsilon
_{0}=\lambda v^{4}/8$.  This vacuum is in an abelian Coulomb phase and the magnetic
field is given by the curl of the Wu-Yang vector potential \cite{wuyang}:%
\begin{align}
&  A_{\varphi}^{N}=g(1-\cos{\theta})\ ,\qquad0\leq\theta\leq\frac{\pi}%
{2}+\epsilon\ ,\nonumber\\
&  A_{\varphi}^{S}=-g(1+\cos{\theta})\ ,\qquad\frac{\pi}{2}-\epsilon\leq
\theta\leq\pi\ .
\end{align}
The matching of the vector bundle gives the Dirac quantization
condition:\footnote{We have already taken in account the fact that the charge
of a fundamental fermion is not $e$ but $e/2$.}%
\begin{equation}
eg=n
\end{equation}
In the external region there is a magnetic field equivalent to the one
generated by a magnetic charge $g$ uniformly distributed on the bag surface:%
\begin{equation}
\vec{B}=g\frac{\hat{r}}{r^{2}}\ ,\qquad r\geq R\ . \label{magprofile}%
\end{equation}
The energy stored in the magnetic field is:%
\begin{equation}
E_{B}=\int\frac{\vec{B}^{2}}{2}=\int_{R}^{\infty}\,4\pi r^{2}\,dr\,\frac
{g^{2}}{2r^{4}}=\frac{2\pi g^{2}}{R}\ . \label{magneticenergy}%
\end{equation}
The $1/R$ dependence means that there is a negative pressure outside the
monopole that tends to expand the bag. To obtain a stable object we need an
opposite force that tends to squeeze the bag. There are three
possible sources for this force: the energy stored in the tail of the scalar
field, the tension of the wall $T_{\mathrm{W}}$ and the internal energy
density $\varepsilon_{0}$.\footnote{This is nothing but the Derrick
\cite{Derrick:1964ww} collapse force coming from the scalar part of the action
$\partial\phi\partial\phi+V(\phi)$.} The radius of the magnetic bag
$R_{\mathrm{m}}$ is determined by the balance between these opposite
oriented\ forces.\footnote{The fact that 't Hooft-Polyakov
monopoles become magnetic bags in the large $n$ limit can also be generalized
to the Julia-Zee dyon \cite{Julia:1975ff}.}

\subsection{Multiple zeros}

We can understand why multi-monopoles become magnetic bags using the analogy
with the multi-vortex case. There are two fundamental reasons why the profile
of the multi-vortex becomes a step function. When $r$ is greater than the
radius of the vortex, the scalar field $\phi$ approaches the vev exponentially%
\begin{equation}
\phi(r)-v\propto e^{-M_{H}r}\ ,\qquad r\gg R\ . \label{grande}%
\end{equation}
Near the center it is instead polynomial
\begin{equation}
\phi(r)\propto r^{n}\ ,\qquad r\ll R \label{piccolo}%
\end{equation}
where $n$ is the number of magnetic fluxes. The only possible way to match
together (\ref{grande}) and (\ref{piccolo}) is a step function in the large
$n$ limit. Now what about the multi-monopole? (\ref{grande}) is still true but
(\ref{piccolo}) must be discussed more carefully. In the multi-vortex case,
(\ref{piccolo}) is a consequence of the fact that the Higgs field winds $n$
times at infinity\ and at $r=0$ must vanish and must be analytic.

\begin{figure}
[tbh]
\begin{center}
\includegraphics[
height=1.7374in,
width=3.7395in
]%
{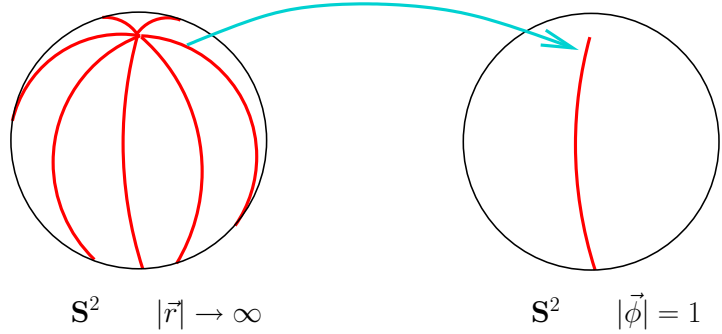}%
\caption{{\protect\footnotesize The axial symmetric map from the sphere
$\mathbf{S}^{2}$  at spatial infinity to the sphere
$\mathbf{S}^{2}$  of the vacuum manifold.}}%
\label{assiale}%
\end{center}
\end{figure}

In the case of the multi-monopole we have to consider the maps from
$\mathbf{S}^{2}$ to $\mathbf{S}^{2}$ and study their analytic behaviour near
zero. The simplest (i.e. the most symmetric) case is the axial symmetric
multi-monopole. The map is represented in Figure \ref{assiale}.
The norm of the Higgs field is $\phi=\sqrt{\phi_{1}^{2}+\phi_{2}^{2}+\phi
_{3}^{2}}$ and the three components have the following behaviour near zero:
\begin{equation}
\phi_{3}\sim z~,\qquad\phi_{1}+i\phi_{2}\sim(x+iy)^{n}~,
\end{equation}
where $n$ is the winding number. Now we take a particular plane passing
through the origin and we want to evaluate the polynomial behavior of the
$\phi$ field near spatial zero. For the axial plane, the plane that is
perpendicular to the axial line, the component $\phi_{3}$ vanishes while the
complex vector $\phi_{1}+i\phi_{2}$ winds $n$ times around zero. This means
that, on the axial plane, the norm of the Higgs field is $\propto r^{n}$ near
the origin and the extra dimension of the bag can be developed. (The magnetic bag is essentially this "fat" zero.)
 If we consider
any other plane that is not the axial one, we have to consider also the
winding of the vectors $\phi_{2}+i\phi_{3}$ and $\phi_{3}+i\phi_{1}$ and then
the norm of the Higgs field vanishes linearly. The result is that in the
$n\rightarrow\infty$ limit we have a magnetic disc that lies in the axial
plane. The thickness of the disc is of order $1/M_{W}$ while its radius is of
order $n/M_{W}$. \ We will analyse in more detail the magnetic disc in Section
\ref{moreonthebag} when we check our statements confronting with the exact
known solution of the axial symmetric multi-monopole.

 For the moment we are interested in a more generic situation where
for every choice of the plane that passes through the origin, the norm of the
Higgs field vanishes\ with some power of order $n$, that is $\phi\propto
r^{\mathcal{O}(n)}$. \ We can also imagine that is possible to find a limit
where there is no preferred direction and the various patches that cover the
$\mathbf{S}^{2}$ sphere are distributed in a \emph{homogeneous}
way.\footnote{This is in fact the case we are going to consider in Section
\ref{monopolesandcosmology} where we address the possible production of
multi-monopoles in the cosmological context.} In this way the multi-monopole
should recover the spherical symmetry at $n\rightarrow\infty$ and should
become a spherical magnetic bag. To see that this is in fact possible we
remand to the discussion on the Jarvis rational map in Subsection \ref{jarvis}.

\subsection{The stabilizing forces and the spectrum of multi-monopoles}

When the potential vanishes, both the tension of the wall and the internal
vacuum energy density are zero. The stabilizing force comes only from the long
tail of the Higgs field $\phi$ outside the monopole. Since the potential is
zero, the mass of the scalar field vanishes and then the profile $\phi(r)$
approaches the vev $v$ like $1/r$ and not exponentially. So the scalar field
profile is:%
\[
|\phi|=0\qquad r\leq R\ ,
\]%
\begin{equation}
|\phi|=v\left(  1-\frac{R}{r}\right)  \qquad r\geq R\ . \label{BPSprofile}%
\end{equation}
The energy coming from this tail is:%
\begin{equation}
\int\frac{1}{2}\partial_{r}\phi\partial_{r}\phi=2\pi v^{2}R\ .
\end{equation}
This brings a pressure force (from outside) that tends to contract the bag and balances with the expansion force coming from the magnetic field
(\ref{magneticenergy}).
\begin{proposition}
The magnetic bag, in the limit of vanishing potential, saturates the
Bogomol'nyi bound.
\end{proposition}
\begin{proof}
The mass bag formula is%
\begin{equation}
M(R)=\frac{2\pi n^{2}}{e^{2}R}+2\pi v^{2}R \label{massbag}%
\end{equation}
minimizing with respect to $R$ we obtain%
\begin{equation}
R_{\mathrm{BPS}}=\frac{n}{ev}\ ,
\end{equation}
and inserting back into (\ref{massbag})
\begin{equation}
M_{\mathrm{BPS}}=\frac{4\pi v}{e}n\
\end{equation}
we obtain exactly the BPS mass.
\end{proof}

Now we introduce a\ tiny potential (\ref{potential}) and the mass bag formula
is%
\begin{equation}
M(R)=\frac{2\pi n^{2}}{e^{2}R}+2\pi v^{2}R+T_{\mathrm{W}}4\pi R^{2}%
+\varepsilon_{0}\frac{4}{3}\pi R^{3}\ , \label{massbagformula}%
\end{equation}
where the tension of the wall is $T_{\mathrm{W}}\sim\sqrt{\lambda}v^{3}$, its
thickness is $\Delta_{\mathrm{W}}\sim1/(\sqrt{\lambda}v)$ and the internal
energy density is $\varepsilon_{0}=\lambda v^{4}/8$. The mass bag formula can
have three different regimes. The first regime is the BPS one where the
tension scales linearly with $n.$ Then we have a SLAC regime where we neglect
the zero energy density $\varepsilon_{0}$ and finally the MIT regime where we
neglect the tension of the wall and consider only the internal energy density.
In the case of a quartic potential such as (\ref{potential}) we have only two
regimes the BPS\ and the MIT ones.\footnote{The previous analysis of the
spectrum can be applied also to a generic potential. The transition between
the BPS and the MIT regimes happens at $n^{\ast}\sim2e\frac{v^{2}}%
{\sqrt{\varepsilon_{0}}}$. The only changes can happen in the case of a
potential where the non-abelian phase $\phi=0$ is strongly metastable. In this
case we should also take in account the presence of a SLAC window between the
two regimes.} So for our purposes
is enough to compute the radius and the mass in the MIT bag regime where we consider only the first and the last terms in
(\ref{massbagformula}). The  minimization gives
\begin{equation}
R_{\mathrm{MIT}}=2^{1/2}\,\frac{1}{e^{1/2}\lambda^{1/4}v}\,~n^{1/2}\ ,\qquad
M_{\mathrm{MIT}}=\frac{2^{5/2}\pi}{3}\,\frac{\lambda^{1/4}v}{e^{3/2}%
}~\,n^{3/2}\ .
\end{equation}

\begin{figure}
[tbh]
\begin{center}
\includegraphics[
height=2.3056in,
width=3.4765in
]%
{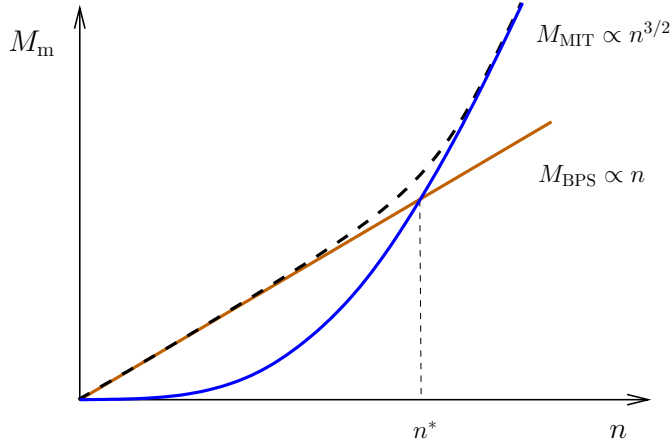}%
\caption{{\protect\footnotesize The spectrum of multi-monopoles (black/dashed
line) is obtained interpolating between the BPS curve (red/solid line) and the
MIT bag curve (blue/solid line). Around $n\sim n^{\ast}$ there is a second
order phase transition between the BPS regime and the MIT regime. }}%
\label{spettro}%
\end{center}
\end{figure}

We are finally ready to analyze the complete spectrum of multi-monopoles. In
Figure \ref{spettro} we show the curves $M_{\mathrm{BPS}}(n)$ and
$M_{\mathrm{MIT}}(n)$ that intersect at%
\begin{equation}
n^{\ast}=\frac{9}{2}\frac{e}{\sqrt{\lambda}}\ .
\end{equation}
Note that in the transition region the radius of the monopole is $R^{\ast}%
\sim1/(\sqrt{\lambda}v)$ that is exactly the inverse of the Higgs boson mass
$1/M_{H}$.
In the BPS region multi-monopoles are marginally stable while in the MIT
region they are instable to decaying into monopoles of lower magnetic charge.
This imply that multi-monopoles are physical objects only if the potential vanishes
or it is very small so that $n^{\ast} \gg 1$.

\section{Moduli Space of BPS Bags \label{modulispace}}

It is a well known fact that BPS solitons admit a moduli space of solutions.
In particular the mass of the $n$-soliton is equal to the sum of its
constituent $1$-soliton masses:%
\begin{equation}
M_{\mathrm{BPS}}(n)=n\,M_{\mathrm{BPS}}(1)\ .
\end{equation}
The aim of this section is to describe the moduli space in the large $n$ limit
of ANO vortices and 't Hooft-Polyakov monopoles.

\subsection{Moduli space of the wall vortex}

\begin{figure}
[ptbh]
\begin{center}
\includegraphics[
height=1.3725in,
width=3.2846in
]%
{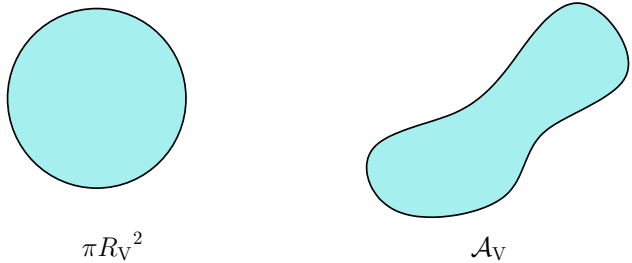}%
\caption{{\protect\footnotesize The moduli space for the BPS wall vortex. Two
closed surfaces with the same area represent two vortices with the same
tension.}}%
\label{area}%
\end{center}
\end{figure}

The moduli space of $n$ BPS vortices, that we indicate by $\mathcal{V}_{n}$,
is a $2n$ real-dimensional space \cite{Weinberg:1979er,Taubes:1979tm}. In the
large $n$ limit multi-vortices become bags and the tension bag formula is%
\begin{equation}
T(R)=\frac{2\pi n^{2}}{e^{2}R^{2}}+\varepsilon_{0}\pi R^{2}\ .
\label{unavariabile}%
\end{equation}
The BPS potential is $V(|q|)=\frac{e^{2}}{2}(|\phi|^{2}-\xi)^{2}$ and the
Coulomb vacuum energy density is thus $\varepsilon_{0}=e^{2}\xi^{2}/2$.
Minimizing (\ref{unavariabile}) with respect to $R$ we obtain exactly the BPS
tension $T_{\mathrm{V}}=2\pi n\xi$. The bag formula (\ref{unavariabile})
refers to a circle of radius $R$. If we substitute the circle with a generic
surface of area $\mathcal{A}$, we obtain%
\begin{equation}
T(R)=\frac{(2\pi n)^{2}}{2e^{2}\mathcal{A}}+\varepsilon_{0}\mathcal{A}\ .
\end{equation}
Any surface that has the same area of the circular wall vortex, that is
$\mathcal{A}_{\mathrm{V}}=\pi(R_{\mathrm{V}})^{2}$, has also the same tension.
The moduli space of wall vortex is thus the set of surfaces in the plane with
fixed area (see Figure \ref{area}).%
Note that, as expected from the large $n$ limit, this is an infinite
dimensional moduli space. It is also easy to see that for the non-BPS vortex
the moduli space is lifted by the contribution of the tension of the wall
$T_{\mathrm{W}}2\pi R$. When the wall tension is added, the minimal tension
for the vortex is obtained when the perimeter is minimized keeping the area
fixed. This implies that the wall vortex of minimal tension is the circular one.

\subsection{Moduli space of the BPS magnetic bag}

Now we return to the main subject of the paper: multi-monopoles and magnetic
bags. The moduli space of $n$ BPS monopoles, that we denote with
$\mathcal{M}_{n}$, is a $4n$ real-dimensional manifold (we are referring to
the $SU(2)$ case) \cite{Weinberg:1979er,Taubes:1979tm,jaffetaubes}. The $4n$
coordinates can be interpreted as the positions of $1$-monopoles plus the
$U(1)$ phase factor.

Now we derive the moduli space of the BPS magnetic bag. We will find that
there are an infinite set of closed surfaces $\mathcal{S}\subset\mathbf{R}%
^{3}$ with the same mass of the spherical surface. It is convenient to
introduce the magnetic scalar potential such that $\vec{B}=\vec{\nabla}%
\varphi$. The Maxwell equation $\vec{\nabla}\cdot\vec{B}=0$ is thus
transformed into the Laplace equation $\Delta\varphi=0$, while the other
Maxwell equation $\vec{\nabla}\wedge\vec{B}=0$ is automatically satisfied. The
magnetic potential outside the bag $\mathcal{S}$ satisfies the Laplace
equation with the following boundary conditions:%
\begin{equation}
\left.  \varphi\right\vert _{\mathcal{S}}=\mathrm{const},\qquad\left.
\int\frac{\partial\varphi}{\partial n}\right\vert _{\mathcal{S}}=4\pi
g\ ,\qquad\left.  \varphi\right\vert _{\infty}=0\ . \label{boundaryone}%
\end{equation}
The scalar field $\phi$ satisfies also the Laplace equation but with different
boundary conditions:%
\begin{equation}
\left.  \phi\right\vert _{\mathcal{S}}=0\ ,\qquad\left.  \phi\right\vert
_{\infty}=v\ . \label{boundarytwo}%
\end{equation}
The mass of the bag is the sum of the energy of the magnetic field plus the
energy of the scalar field:%
\begin{equation}
M(\mathcal{S})=\int\frac{(\vec{\nabla}\varphi)^{2}}{2}+\int\frac{(\vec{\nabla
}\phi)^{2}}{2}%
\end{equation}

Let us recall for a moment the case of a spherical bag. If we denote $R$ the
radius of the sphere, the potential $\varphi$ and the scalar field $\phi$ are
given by%
\begin{equation}
\varphi=-\frac{g}{r}\ ,\qquad\phi=v\left(  1-\frac{R}{r}\right)  \ .
\end{equation}
The mass as function of the radius is%
\begin{equation}
M(R)=\frac{2\pi g^{2}}{R}+2\pi v^{2}R\ ,
\end{equation}
and the minimization gives%
\begin{equation}
R_{\mathrm{m}}=\frac{g}{v}\ ,\qquad m_{\mathrm{m}}=4\pi gv\ .
\label{sphericalbagmass}%
\end{equation}
\begin{proposition}
There is an infinite set of bags with different shapes that saturates the
Bogomol'nyi bound.
\end{proposition}
\begin{proof}
To construct the generic magnetic bag we need the following trick. Consider a
generic function from the ball $\mathbf{B}^{3}$ of radius $1$ to the space
$\mathbf{R}^{3}$:%
\begin{equation}
f:y\in\mathbf{B}^{3}\longrightarrow\mathbf{R}^{3}%
\end{equation}
and then solve the Laplace equation with a source given by the image of the
function $\vec{f}$:%
\begin{equation}
\Delta\varphi(x)=4\pi g\int_{\mathcal{B}^{3}}\delta^{(3)}\left(
x-f(y)\right)  ~\det\left(  \frac{\partial f}{\partial y}\right)  ~d^{3}y
\label{laplace}%
\end{equation}
Physically we can think of it in this way. We have a magnetic source of charge
$g$ that is distributed on a compact region $f(\mathcal{B}^{3})$ with generic
shape and generic distribution of magnetic charge. In Figure \ref{ball} (a) we
have the blob and the surfaces of constant $\phi$. In the following we will see
that a particular surface of this set give rise to a magnetic bag with the same mass as that\ of the
spherical bag.%
\begin{figure}
[tbh]
\begin{center}
\includegraphics[
height=1.6509in,
width=4.4477in
]%
{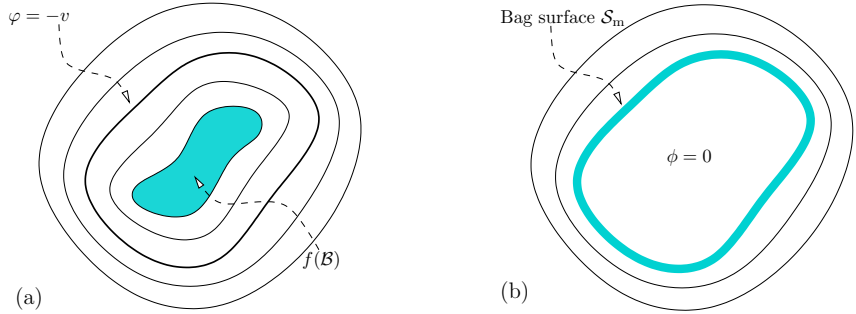}%
\caption{{\protect\footnotesize \underline{(a)}: $f({\bf B}^3)$ is a compact source of magnetic field.
The closed surface are the ones where the magnetic scalar potential $\varphi$
is constant. For $f({\bf B}^3)$ sufficiently compact, there will exist a particular
surface where the value of the potential $\varphi$ is
equal to $-v$. \underline{(b)}: The magnetic bag is obtained taking the surface given by the previous
construction and then distributing on it a magnetic charge equivalent to the
one in the blob of (a). (We stress that (a) is the mathematical construction while (b) is the physical result.)}}%
\label{ball}%
\end{center}
\end{figure}
Consider one of the surfaces of the Figure \ref{ball} (a). We want a magnetic bag
where $\varphi$ is given by the solution of (\ref{laplace}) outside this
surface. Note that this automatically satisfies the boundary conditions
(\ref{boundaryone}). If we want to minimize the energy we have to satisfy the
abelian BPS equations
\begin{equation}
\vec{\nabla}\varphi=\vec{\nabla}\phi~. \label{abelianBPSequation}%
\end{equation}
To do so we simply have to choose the particular surface where%
\begin{equation}
\left.  \varphi\right\vert _{\mathcal{S}_{\mathrm{m}}}=-v\ . \label{condition}%
\end{equation}
This is the central point of the construction. It is easy to see that the
following choice for the Higgs field%
\begin{equation}
\phi=v+\varphi
\end{equation}
satisfies the Laplace equation and the boundary conditions (\ref{boundarytwo})
and also the abelian BPS equation (\ref{abelianBPSequation}).

A check that the energy saturates the Bogomol'nyi bound  is given by a simple
application of the Green's first identity \cite{jackson}%
\begin{equation}
\int_{V}\frac{(\vec{\nabla}\varphi)^{2}}{2}=-\frac{1}{2}\int_{\mathcal{S}%
}\varphi\frac{\partial\varphi}{\partial\vec{n}}=-2\pi g\left.  \varphi
\right\vert _{\mathcal{S}}\ .
\end{equation}
and so
\begin{equation}
\int_{V}\frac{(\vec{\nabla}\varphi)^{2}}{2}=\int_{V}\frac{(\vec{\nabla}%
\phi)^{2}}{2}=2\pi gv\ . \label{equal}%
\end{equation}
The sum of the two is $4\pi gv$ which is exactly the Bogomol'nyi bound.
\end{proof}

The knowledge of the moduli spaces $\mathcal{V}_{n}$ and $\mathcal{M}_{n}$ is
essential to describe the low energy dynamics of vortices and monopoles. The
motion of these particles is described by geodesics in the moduli space
\cite{Manton:1977er}. It is thus fundamental to know not only the topology but
also the metric of these spaces. For the monopoles the situation is better
since we know that $\mathcal{M}_{n}$ is an hyper-K\"{a}hler manifold. This
enabled Atiyah and Hitchin to find the exact metric of the moduli space of two
monopoles and consequently to describe their scattering \cite{Atiyah:1985dv}.
The same process for vortices can be described only using numerical
computations \cite{Samols:1991ne,Myers:1991yh}. A method of Manton permits one
to obtain by simple arguments the metric of monopoles and vortices when they
are far apart \cite{Manton:2002wb}. It is interesting to note that our result
shed light on a complete opposite regime, when a large number of solitons are
very close to each others.

We want to make another conclusive remark to this section. The large $n$ limit
we are studying in this paper can be seen as a sort of linearization of the
problem. The \textquotedblleft non-abelianity\textquotedblright\ is restricted
only to a small shell of thickness negligible compared to the radius of the
bag. The essential parameter, such as the radius $R_{\mathrm{m}}$ and the mass
$M_{\mathrm{m}}$ of the multi-monopoles, can be computed only using the
abelian theory outside the bag. This suggest an intriguing relation with the
linearization of the Nahm equations in the $n\rightarrow\infty$ limit found in
\cite{wardlinearization}.

\section{More on the Magnetic Bag\label{moreonthebag}}

This section has a double aim. First we want to clarify better the magnetic
bag conjecture in the case of the BPS monopole. Second we want to test the
conjecture using the Ward-Prasad-Rossi (WPR) \cite{Ward:1981jb}\cite{prasad}
solution for the axial symmetric multi-monopole.

For simplicity we work in units where $e=v=1$ and the Lagrangian is%
\begin{equation}
\mathcal{L}=-\frac{1}{4}F_{\mu\nu}^{a}F^{\mu\nu a}-\frac{1}{2}D_{\mu}\phi
^{a}D^{\mu}\phi^{a}\ .\label{firstorder}%
\end{equation}
The non-abelian BPS equations are%
\begin{equation}
B_{i}^{a}=\pm D_{i}\phi^{a}%
\end{equation}
where $B_{i}^{a}=\frac{1}{2}\epsilon_{ijk}F_{jk}^{a}$ is the non-abelian
magnetic field. The 't Hooft-Polyakov monopole has the following structure%
\begin{equation}
\phi^{a}=\hat{r}^{a}H(r)~,\qquad A_{i}^{a}=-\epsilon_{aij}\frac{\hat{r}^{j}%
}{r}G(r)
\end{equation}
and the profile functions are%
\begin{equation}
H(r)=\coth r-\frac{1}{r}~,\qquad G(r)=1-\frac{r}{\sinh r}~.
\end{equation}
Using the identity $\partial_{i}\hat{r}_{j}=\left(  \delta_{ij}-\hat{r}%
_{i}\hat{r}_{j}\right)  /r$ we can compute the non-abelian magnetic field
\begin{equation}
B_{i}^{a}=\underset{\mathrm{Abelian}}{\underbrace{-\hat{r}^{a}\hat{r}%
^{i}H^{\prime}(r)}}+\underset{\mathrm{Non-Abelian}}{\underbrace{\left(
\delta_{ia}-\hat{r}^{a}\hat{r}^{i}\right)  \frac{H(r)}{\sinh r}}}~.
\end{equation}
The first piece is the magnetic field projected in the $\phi(r)$ direction
(note that it vanishes like $1/r^{2}$ at infinity) and thus corresponds to the
abelian magnetic field of the unbroken $U(1)$. The second term is a pure
non-abelian field and vanishes exponentially at infinity.

\begin{figure}
[tbh]
\begin{center}
\includegraphics[
height=2.4085in,
width=3.6054in
]%
{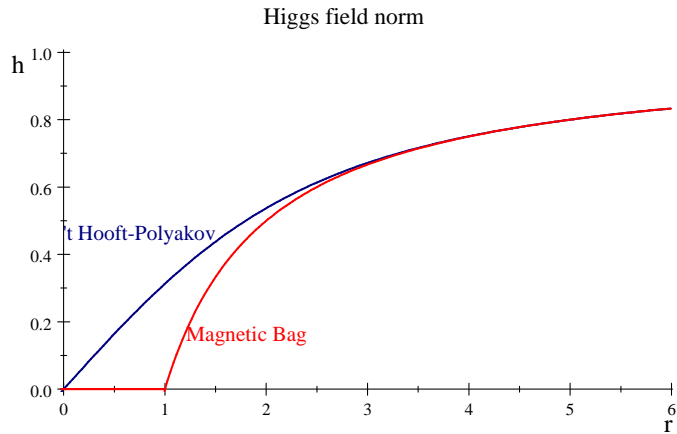}%
\caption{{\protect\footnotesize The norm of the Higgs field for the
$n=1$  monopole and for the
$n=\infty$ monopole. The
radius of the multi-monopole is kept fixed while making the large
$n$ limit.}}%
\label{normhiggs}%
\end{center}
\end{figure}

\begin{figure}
[tbh]
\begin{center}
\includegraphics[
height=2.4094in,
width=3.6054in
]%
{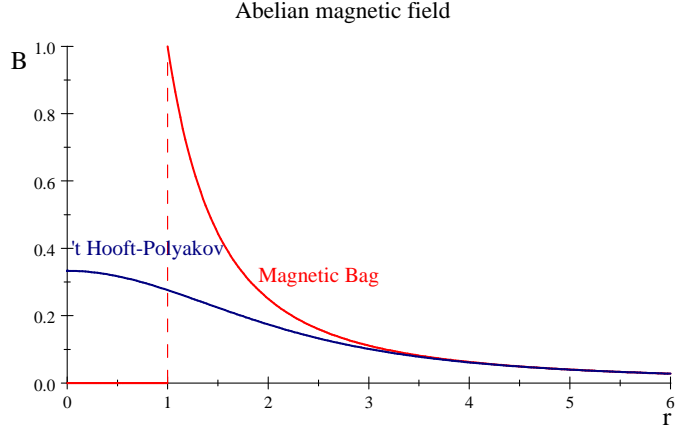}%
\caption{{\protect\footnotesize The abelian projection of the magnetic
field for the $n=1$  monopoles
and for the $ n=\infty$ monopole. The radius of the multi-monopole is kept fixed while making the
large $ n$  limit.}}%
\label{babelian}%
\end{center}
\end{figure}

Now suppose that we have all the multi-monopole solutions that we want, how we
can check the magnetic bag conjecture? The idea is simple and it is the same
we have used for the multi-vortex in \cite{proof}. While we are making the
large $n$ limit we also rescale the lengths such that the radius of the
monopole remains constant $r\rightarrow r/n$. In this case, if the magnetic
bag conjecture is true, it would appear very clearly as in Figure
\ref{normhiggs} for the norm of the Higgs field and Figure \ref{babelian} for
the abelian part of the magnetic field.%

Unfortunately not all the multi-monopole solutions are known. But one solution
is known in great detail: the axial symmetric multi-monopole. In what follows
we are going to confront our conjecture with this solution and we will find an
encouraging agreement with the expectations.

\subsection{The axial symmetric multi-monopole}

The WPR axial symmetric multi-monopole is an exact solution for monopoles of
multiple magnetic charge. The axial symmetry can be seen in Figure
\ref{assiale} where we have the map from the sphere $\mathbf{S}^{2}$ at
spatial infinity to the sphere $\mathbf{S}^{2}$ of the vacuum manifold
$|\vec{\phi}|=1$. The spatial sphere covers the vacuum manifold sphere $n$
times winding around the $\hat{z}$ axis.

The $n=1$ case corresponds to the Prasad-Sommerfield \cite{Prasad:1975kr}
solution while the $n=2$ case is the multi-monopole of charge $2$ first
obtained in \cite{Ward:1981jb}. The generalization to arbitrary $n$ has been
found in \cite{prasad}. The axial symmetric multi-monopole corresponds to one
particular point in the moduli space of multi-monopoles. Say in another way,
the imposition of the axial symmetry fixes all the degrees of freedom in the
moduli space, apart from a global translation and rotation.

\begin{figure}
[tbh]
\begin{center}
\includegraphics[
height=3.0753in,
width=3.7395in
]%
{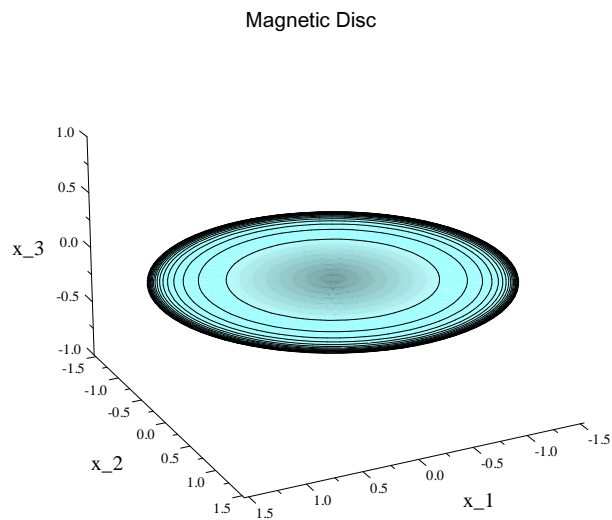}%
\caption{{\protect\footnotesize The magnetic disc centered in zero, with
radius $\frac{\pi}{2}$  and with the axial line
oriented in the $x_{3}$ direction. The density of the concentric lines is proportional to the density
of magnetic charge. }}%
\label{disc}%
\end{center}
\end{figure}

Now let's find the exact value of the radius of the magnetic disc. To find it
we must repeat the procedure we have done for the magnetic sphere. We take a
generic magnetic disc of radius $R_{\mathrm{d}}$ with magnetic charge $1$. The
Higgs field $\phi$ is a solution to the Laplace equation with Dirichlet
boundary conditions: $\phi=0$ on the disc and $\phi=1$ at infinity. The
solution can be written in cylindrical coordinates using the expansion in
Legendre polynomials \cite{jackson}:%
\begin{equation}
\phi(r,\theta)=\left\{
\begin{array}
[c]{cc}%
1-\frac{2}{\pi}\frac{R_{\mathrm{d}}}{r}\sum_{l=0}^{\infty}\frac{\left(
-1\right)  ^{l}}{2l+1}\left(  \frac{R_{\mathrm{d}}}{r}\right)  ^{2l}%
P_{2l}\left(  \cos\theta\right)  ~,\qquad & r\geq R_{\mathrm{d}}\\
& \\
\frac{2}{\pi}\sum_{l=0}^{\infty}\frac{\left(  -1\right)  ^{l}}{2l+1}\left(
\frac{r}{R_{\mathrm{d}}}\right)  ^{2l+1}P_{2l+1}\left(  \cos\theta\right)
~,\qquad & r\leq R_{\mathrm{d}}%
\end{array}
\right.  \label{series}%
\end{equation}
The magnetic scalar potential $\varphi$ is again a solution to the Laplace
equation but with different boundary conditions: $\varphi=0$ at infinity and
$\varphi=\mathrm{const}$ on the disc, where the constant is fixed imposing
that the charge of the disc is $1$. The solution is%
\begin{equation}
\varphi\left(  r,\theta\right)  =\left\{
\begin{array}
[c]{cc}%
\frac{1}{r}\sum_{l=0}^{\infty}\frac{\left(  -1\right)  ^{l}}{2l+1}\left(
\frac{R_{\mathrm{d}}}{r}\right)  ^{2l}P_{2l}\left(  \cos\theta\right)  ,\qquad
& r\geq R_{\mathrm{d}}\\
& \\
\frac{\pi}{2R_{\mathrm{d}}}-\frac{1}{R_{\mathrm{d}}}\sum_{l=0}^{\infty}%
\frac{\left(  -1\right)  ^{l}}{2l+1}\left(  \frac{r}{R_{\mathrm{d}}}\right)
^{2l+1}P_{2l+1}\left(  \cos\theta\right)  ,\qquad & r\leq R_{\mathrm{d}}%
\end{array}
\right.
\end{equation}
where it turns out that$\ \varphi=\frac{\pi}{2R_{\mathrm{d}}}$ on the disc. To
obtain the radius of the disc we must sum the energy carried by the two scalar
potentials and then minimize with respect to $R_{\mathrm{d}}$ (the same we
have done in (\ref{massbag}) for the spherical bag). But we can use a shortcut
since we know that the energy is minimized exactly when the two contribution
are equal (see the abelian BPS equation (\ref{abelianBPSequation})), and we easily obtain the radius of the disc:%
\begin{equation}
R_{\mathrm{d}}=\frac{\pi}{2}~.
\end{equation}
The magnetic charge on the disc has distribution $\frac{1}{\pi^{2}}\frac
{1}{\sqrt{\left(  \frac{\pi}{2}\right)  ^{2}-\left(  x_{1}^{2}+x_{2}%
^{2}\right)  }}$. The magnetic disc (Figure \ref{disc}) is consistent with
what we discussed in Section \ref{modulispace}. The shape of the bag is now
degenerate and squeezed in the $x_{3}$ direction.

\begin{figure}
[tbh]
\begin{center}
\includegraphics[
height=2.7216in,
width=4.0888in
]%
{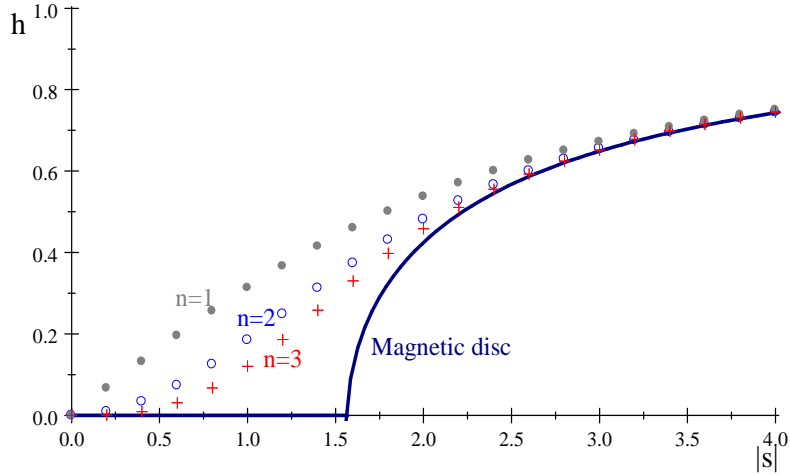}%
\caption{{\protect\footnotesize We have plotted the norm of the Higgs field on
the axial plane as function of the radius. The dots, circles and crosses
refers respectively to $ n=1,2,3$, and have been computed using (\ref{formuladue}) with
the radius rescaled by a factor of $n$.
The line is the magnetic disc potential and has been
computed from (\ref{series}). }}%
\label{axialplane}%
\end{center}
\end{figure}

\begin{figure}
[tbh]
\begin{center}
\includegraphics[
height=2.7077in,
width=4.0542in
]%
{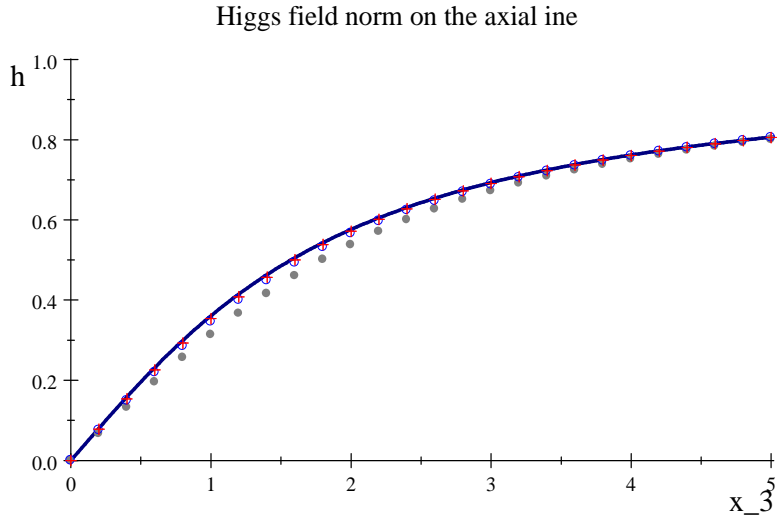}%
\caption{{\protect\footnotesize We have plotted the norm of the Higgs field on
the axial line as function of the $x_{3}$ coordinate. The conventions are the same of Figure
\ref{axialplane}.}}%
\label{axialline}%
\end{center}
\end{figure}

Now it is time to confront with the WPR solution. We will not give the details
of the derivation, the reader can find them in the literature (in particular
in Ref. \cite{prasad}). We will just present the formulas that are needed for
the solution. First of all we have to introduce the functions $\Delta
_{n,l}(x_{i=1\dots4})$%
\begin{equation}
\Delta_{n,l}(x_{i=1\dots4})=\frac{1}{2}(-1)^{l}e^{ix_{4}-il\theta}\int
_{-1}^{1}\left[  2\cos\left(  \frac{\pi t}{2}\right)  \right]  ^{n-1}%
e^{-x_{3}t}\left(  \frac{1+t}{1-t}\right)  ^{l/2}I_{l}\left(  s\sqrt{1-t^{2}%
}\right)  dt
\end{equation}
where $n$ is the winding number, $l$ is an integer that goes from $-n$ to $n$,
$x_{1,2,3}$ are the spatial coordinates, $s$ is the complex variable
$x_{1}+ix_{2}$ and $x_{4}$ is the Euclidean time. We can thus introduce the
three potentials $\phi_{n}$,$\rho_{n}$ and $\chi_{n}$ that are needed for the
construction:%
\begin{equation}
\phi_{n}\left(  x_{i=1\dots4}\right)  =(-1)^{n+1}\frac{\det\left(
\begin{array}
[c]{cccc}%
\Delta_{n,-n+1} & \cdots & \cdots & \Delta_{n,0}\\
\vdots &  &  & \vdots\\
&  &  & \\
\Delta_{n,0} & \cdots & \cdots & \Delta_{n,n-1}%
\end{array}
\right)  }{\det\left(
\begin{array}
[c]{ccc}%
\Delta_{n,-n+2} & \cdots & \Delta_{n,0}\\
\vdots &  & \vdots\\
\Delta_{n,0} & \cdots & \Delta_{n,n-2}%
\end{array}
\right)  }~,
\end{equation}%
\begin{equation}
\rho_{n}\left(  x_{i=1\dots4}\right)  =(-1)\frac{\det\left(
\begin{array}
[c]{cccc}%
\Delta_{n,-n} & \cdots & \cdots & \Delta_{n,-1}\\
\vdots &  &  & \vdots\\
&  &  & \\
\Delta_{n,-1} & \cdots & \cdots & \Delta_{n,n-2}%
\end{array}
\right)  }{\det\left(
\begin{array}
[c]{ccc}%
\Delta_{n,-n+2} & \cdots & \Delta_{n,0}\\
\vdots &  & \vdots\\
\Delta_{n,0} & \cdots & \Delta_{n,n-2}%
\end{array}
\right)  }~,
\end{equation}%
\begin{equation}
\chi_{n}\left(  x_{i=1\dots4}\right)  =(+1)\frac{\det\left(
\begin{array}
[c]{cccc}%
\Delta_{n,-n+2} & \cdots & \cdots & \Delta_{n,1}\\
\vdots &  &  & \vdots\\
&  &  & \\
\Delta_{n,1} & \cdots & \cdots & \Delta_{n,n}%
\end{array}
\right)  }{\det\left(
\begin{array}
[c]{ccc}%
\Delta_{n,-n+2} & \cdots & \Delta_{n,0}\\
\vdots &  & \vdots\\
\Delta_{n,0} & \cdots & \Delta_{n,n-2}%
\end{array}
\right)  }~.
\end{equation}
In terms of these potentials it is possible to write the solution of the
non-abelian BPS equations (that is the potentials $A_{i}$ and $\phi$). The
problem is that this solution is expressed in a complex gauge and, even if it
is proven to be gauge equivalent to a real one, the explicit form for the real
solution is not known. Fortunately there exist a simple expression for the
modulus of the Higgs field%
\begin{equation}
h_{n}\left(  r,x_{3}\right)  =\left\vert \frac{1}{\phi_{n}}\sqrt{\left(
\frac{\partial\phi_{n}}{\partial x_{3}}\right)  ^{2}-\left(  \rho_{n}%
-\frac{\partial\rho_{n}}{\partial x_{3}}\right)  \left(  \chi_{n}%
+\frac{\partial\chi_{n}}{\partial x_{3}}\right)  }\right\vert ~. \label{norm}%
\end{equation}
This expression gives the modulus of the Higgs field as function of the radius
$r=\sqrt{x_{1}^{2}+x_{2}^{2}}$ and the coordinate $x_{3}$. There are two
convenient simplifications of this formula. On the axial plane (\ref{norm})
reduces to
\begin{equation}
h_{n}\left(  r,0\right)  =\left\vert \frac{1}{\phi_{n}}\left(  \rho_{n}%
-\frac{\partial\rho_{n}}{\partial x_{3}}\right)  \right\vert _{x_{3}%
=0}=\left\vert \frac{1}{\phi_{n}}\left(  \chi_{n}+\frac{\partial\chi_{n}%
}{\partial x_{3}}\right)  \right\vert _{x_{3}=0}~. \label{formuladue}%
\end{equation}
On the axial line (\ref{norm}) reduces to%
\begin{equation}
h_{n}\left(  0,x_{3}\right)  =\left\vert \partial_{x_{3}}\ln\Delta
_{n,l}(0,0,x_{3},0)\right\vert ~. \label{formulauno}%
\end{equation}
Now we are ready for the final step: the computation. Using a numerical
computation\footnote{We have used the program Scientific WorkPlace.} we have
been able to find the norm of the Higgs field up to $n=3$. The hardest step
for the computation is the derivative with respect to $x_{3}$, and for this
reason we are able only to go to $n=3$. Anyway the results give a quite
convincing proof of the conjecture. In Figure \ref{axialplane} we present the
norm of the Higgs field on the plane perpendicular to the axial line using the
formula (\ref{formuladue})\ while in Figure \ref{axialline} we present the
norm on the axial line using the formula (\ref{formulauno}). The points in the
plots correspond to the norm for $n=1,2,3$ (respectively green/dot,
blue/circle and red/cross). The gray line is instead the norm of the Higgs
field for the magnetic disc computed using the formula (\ref{series}) with
$R_{\mathrm{d}}=\frac{\pi}{2}$.%
On the axial line $r=0$ the convergence can be expressed in a compact
mathematical form%
\begin{equation}
\lim_{n\rightarrow\infty}\frac{\int_{-1}^{1}ze^{-ntx_{3}}\cos^{n-1}\frac{\pi
t}{2}dt}{\int_{-1}^{1}e^{-ntx_{3}}\cos^{n-1}\frac{\pi t}{2}dt}=\frac{2}{\pi
}\arctan\frac{2x_{3}}{\pi}~, \label{greataccuracy}%
\end{equation}
where the right hand side simply is (\ref{formulauno}) and the left hand side
is the resummation of the series (\ref{series}). Formula (\ref{greataccuracy})
can be verified with great accuracy\ but we don't know a simple analytic proof
of it.\footnote{In the Figure \ref{axialline} we have showed only the values
$n=1,2,3$ but formula (\ref{greataccuracy}) can be verified \ much beyond. On
the axial plane (Figure \ref{axialplane}) $n=3$ is instead our computational
limit.}

\subsection{Large $n$ limit of multi-monopoles \label{jarvis}}

To understand completely the large $n$ limit of multi-monopoles, we have to
understand in which way the position in the moduli space is related to the
shape of the magnetic bag. In this paper we have provided two examples. In the
first one we choose a limit in which the spherical symmetry is asymptotically
restored in the large $n$ limit and the outcome is a bag with spherical shape and
radius $n$ (in units $e=v=1$). In the second example we have the axial
symmetric multi-monopole. The outcome is a bag with a degenerate shape: a disc
with radius $\frac{n\pi}{2}$ perpendicular to the axial line of the multi-monopole.

Among the various mathematical approaches to the study of multi-monopoles, the
Jarvis \cite{Jarvis} rational map is probably the most suitable to
the large $n$ limit problem.\footnote{The Jarvis rational map is a modified
version of the Donaldson rational map \cite{Donaldson:1985id}. In the
Donaldson we have to choose a particular direction in $\mathbf{R}^{3}$ and
this break the rotational invariance. In the Jarvis case we have to choose an
origin. This preserves the rotational invariance around the origin.} In the
Jarvis framework we have to choose an origin in the space $\mathbf{R}^{3}$ and
then consider the Hitchin equation $(D_{r}-i\Phi)s=0$ along each radial line
from the origin to infinity. $s$ is a doublet in the $SU(2)$ representation
and there is only one solution which decays asymptotically as $r\rightarrow
\infty.$ Call this solution $\left(
\begin{array}
[c]{c}%
s_{1}(r)\\
s_{2}(r)
\end{array}
\right)  $ and $\left(
\begin{array}
[c]{c}%
s_{1}(0)\\
s_{2}(0)
\end{array}
\right)  $ its value at the origin. The Jarvis rational map is given by
$R=\frac{s_{1}(0)}{s_{2}(0)}$ and is a correspondence between Riemann\ spheres
$R:\mathbf{CP}^{1}\mapsto\mathbf{CP}^{1}$. \ The map is holomorphic since, due
to the Bogomol'nyi equation, the operator $D_{r}-i\Phi$ commutes with
$D_{\bar{z}}$. A gauge transformation replace $R$ by a $SU(2)$ Mobius
transformation determined by the gauge transformation at the origin. We thus
have a correspondence between the moduli space of the $n$-monopole
and the equivalence class of unbased rational maps of degree $n$.\footnote{The rational map has $4n+2$ parameters. Subtracting the $3$ parameters
of the $SU(2)$ gauge transformation gives $4n-1$. This is the dimension of gauge inequivalent n-monopoles. In general a $U(1)$ phase is then
added for convenience. }

A procedure to obtain the monopole solution from the Jarvis rational map has
been studied in \cite{Ioannidou:1999iw} and applied to the study of
particularly symmetric multi-monopoles. Choosing a certain complex gauge, the
boundary conditions for the field $\phi$ and $A_{i}$ can be expressed in terms
of the rational map. In particular the map from the sphere $\mathbf{S}^{2}$ at
spatial infinity to the sphere $\mathbf{S}^{2}$ of the vacuum manifold is
essentially given by the rational map itself.

Using the Jarvis rational map we can at least speculate what could be a large
$n$ limit for multi-monopoles. The map can be thought of as a certain distribution
of $n$ zeros and $n$ poles over the Riemann sphere. When $n$ is large the
zeros and poles form a continuous distribution. Call $\sigma_{0}(z)$ and
$\sigma_{\infty}(z)$ the density of zeros and poles over the sphere divided by
the total number $n$. We can define a large $n$ limit so that the densities
$\sigma_{0}(z)$ and $\sigma_{\infty}(z)$ remain constant. For a homogeneous
distribution $\sigma_{0}(z)=\sigma_{\infty}(z)=\frac{1}{4\pi}$ and we should
obtain the spherical magnetic bag. The rational map for the axial symmetric
multi-monopole is $R(z)=z^{n}$ and is a highly degenerate distribution with
$n$ zeros in zeros and $n$ poles at infinity. It is an open question what is
the magnetic bag surface corresponding to generic distributions $\sigma
_{0}(z)$ and $\sigma_{\infty}(z)$.

\section{String Theory Interpretation of Bag Solitons\label{string}}

In this section we will to interpret our result in the string theory context.
For the following results we refer in particular to the reviews
\cite{Giveon:1998sr} for brane setups of gauge theories and \cite{tongreview}
for solitons.

\begin{figure}
[tbh]
\begin{center}
\includegraphics[
height=1.8983in,
width=3.7706in
]%
{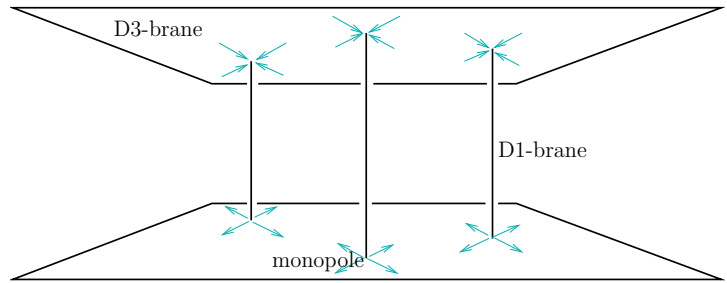}%
\caption{{\protect\footnotesize The $\mathcal{N}=4$ $U(2)$ gauge theory is
realized on the world volume of two D$3$-branes in type IIB string theory.
D$1$-branes stretched between the two D$3$-branes correspond to BPS monopoles
in the four dimensional gauge theory.}}%
\label{duebstandard}%
\end{center}
\end{figure}

\begin{figure}
[tbh]
\begin{center}
\includegraphics[
height=1.9in,
width=3.7715in
]%
{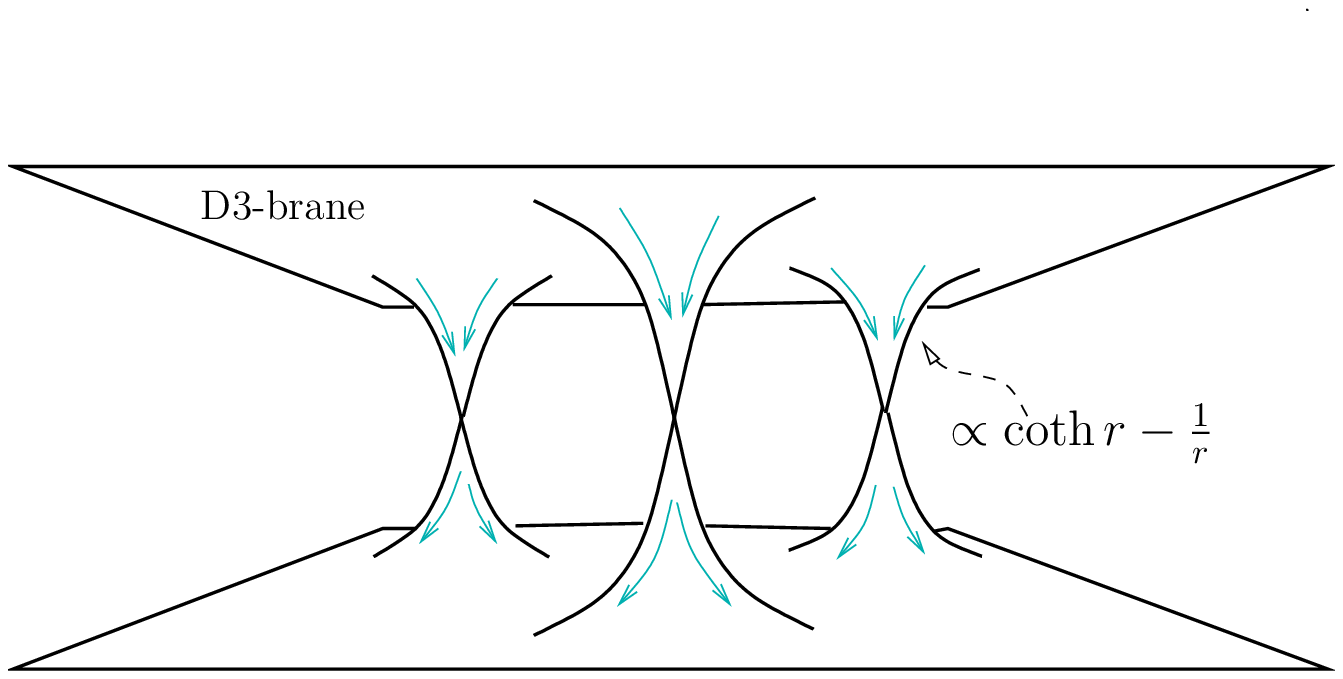}%
\caption{{\protect\footnotesize D$1$-branes suspended
between two D$3$-branes as seen from the DBI\ action
point of view. }}%
\label{duebhashimoto}%
\end{center}
\end{figure}

Monopoles in string theory can be obtained in the following way. We take a
stack of $N$ D$3$-branes in type IIB string theory. The low energy theory that
describes the dynamics of the branes is a $\mathcal{N}=4$ $U(N)$ gauge theory
in $3+1$ dimensions. In Figure \ref{duebstandard} we have a $U(2)$ gauge
theory broken down to $U(1)\times U(1)$. Monopoles correspond to D$1$-branes
stretched between the D$3$-branes; the point where they end on the brane is
the position of the monopole.
The brane setup of Figure \ref{duebstandard} is just a classical cartoon. When
D$1$-branes end on a D$3$-brane they create a disturbance in its shape due to
their tension. Figure \ref{duebstandard} is a reliable approximation only when
the distance between the D$3$-branes is sufficiently large and the monopoles
are far enough away from each other so that the disturbances do not overlap. A
way to study the disturbance created by a D$1$-brane ending on a D$3$-brane
has been developed in \cite{Bions}. The effective theory that describes the
low energy degrees of freedom on a single D$3$-brane is the Dirac-Born-Infeld
(DBI) theory. This non-linear theory possesses non-trivial solutions to the
classical equation of motion that can be identified with the D$1$-brane ending
on the D$3$-brane. This solution, also called a BIon, is a spike coming out of
the D$3$-brane with a profile proportional to $1/r$. This shows that the
D$1$-brane is made of the same substance as that of the D$3$-brane. When
considering D$1$-branes suspended between two D$3$-branes we need to use the
non-abelian generalization of the DBI action. \ It has been shown in
\cite{hashimoto} that for BPS quantities the solution of the non-abelian DBI
action is also a solution to its first order expansion (\ref{firstorder}%
).\footnote{Still there is an uncertainty about the way to take the traces in
the non-abelian DBI action. Ref. \cite{hashimoto} uses the prescription given
in Ref. \cite{nonabelianprescription}.} The suspended D$1$-brane is thus
described by two spikes of D$3$-branes with profile proportional to $\coth
r-\frac{1}{r}$ that meet in a single point. Figure \ref{duebhashimoto} is what
happens to the D$1$-branes when their distance is large enough for the spikes
not to be overlapped.
In the magnetic bag limit we are in the opposite regime where the disturbances
created by the D$1$-branes are completely overlapped so it no more is
meaningful to speak about the position of the single D$1$-brane. What emerges
is instead the magnetic bag surface $\mathcal{S}_{\mathrm{m}}$ of radius
$\propto n$ where the two spikes coming out from the D$3$-branes are joined
together (see Figure \ref{magneticbag}). The two spikes have profiles
proportional to $1-\frac{n}{r}$. In the interior of the magnetic bag the two
D$3$-branes are on the top of each other.%

\begin{figure}
[tbh]
\begin{center}
\includegraphics[
height=1.887in,
width=3.7611in
]%
{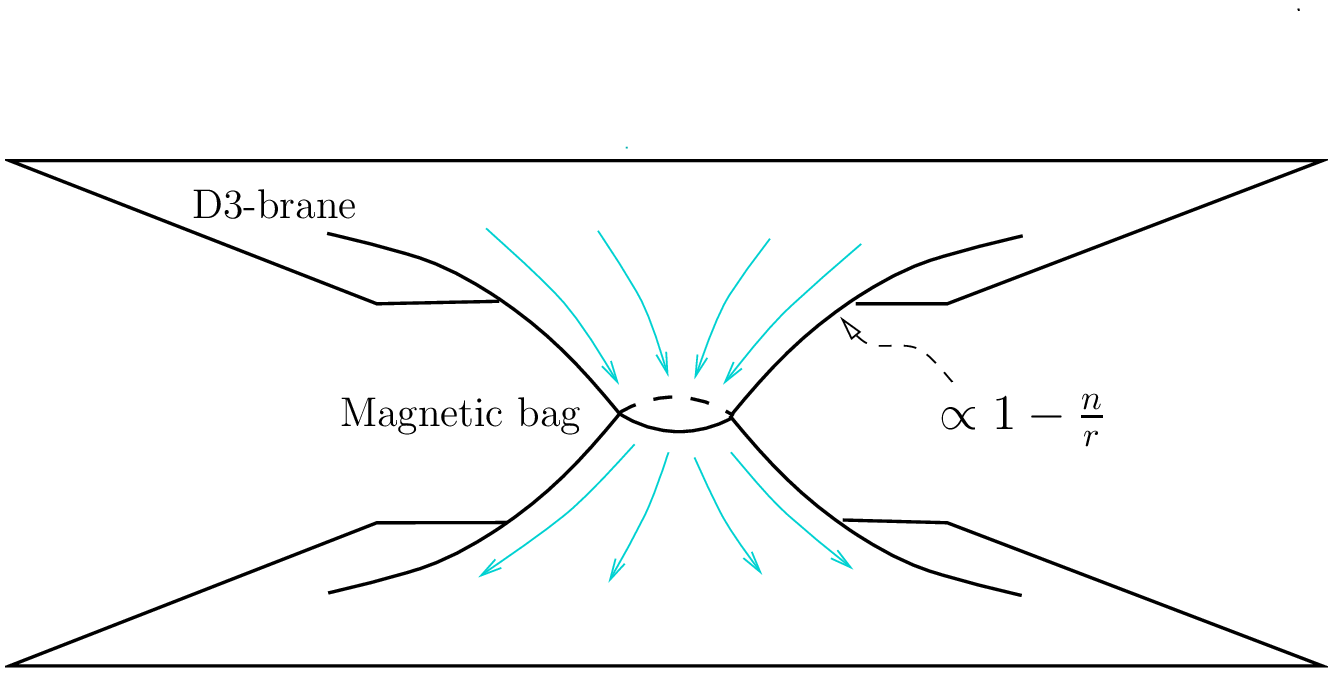}%
\caption{{\protect\footnotesize The magnetic bag is a tube of D$3$-branes
connecting the two gauge $3$-branes. The magnetic flux passes trough the tube and
stabilizes it. }}%
\label{duebsacca}%
\end{center}
\end{figure}

Now we consider wall vortices in the string theory context. Even in this case
a \textquotedblleft brane transmutation\textquotedblright\ effect will be the
string theoretical explanation. We will see that by T-duality the wall vortex
is essentially the same object as the magnetic bag.

The four dimensional theory is $\mathcal{N}=2$ SYM with gauge group $U(N_{c})$
and $N_{f}$ matter hypermultiplets with masses $m_{i}$. This theory is broken
to $\mathcal{N}=1$ by a superpotential $\mathcal{W}(\Phi)$ for the adjoint
chiral superfield. The moduli space of the $\mathcal{N}=2$ theory is lifted
and only a discrete number of vacua survive \cite{Cachazo:2003yc}. We are
interested in vacua where some diagonal element of the adjoint scalar field
$\phi_{j}$ is equal to some flavor mass $m_{i}$. Vortices arise in the
color-flavor locked vacua.

\begin{figure}
[tbh]
\begin{center}
\includegraphics[
height=2.3618in,
width=3.7836in
]%
{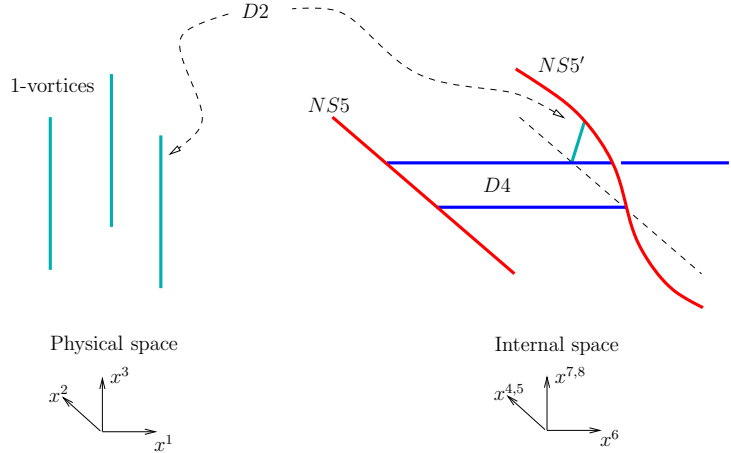}%
\caption{{\protect\footnotesize Brane setup for the $\mathcal{N}=2$ theory
broken down\ to $\mathcal{N}=1$ by a superpotential. Vortices in the physical
space correspond to D$2$-branes stretched between the color-flavor locked
D$4$-brane and the NS$5^{\prime}$-brane.}}%
\label{cartoonstandard}%
\end{center}
\end{figure}

The brane realization is obtained in type IIA string theory as follows. The
$\mathcal{N}=2$ theory is obtained with two NS$5$-branes extended in
$x^{0,1,2,3,4,5}$ at the positions $x^{6}=0$ and $x^{6}=L$ (we call them NS$5$
and NS$5^{\prime}$). Then there is a stack of $N_{c}$ D$4$-branes extended in
$x^{0,1,2,3,6}$ between the two NS$5$-branes. Finally there is a set of
$N_{f}$ semi-infinite D$4$-branes that end on the NS$5^{\prime}$-brane
\cite{Witten:1997sc}. The breaking to $\mathcal{N}=1$ is obtained by giving a
shape to the NS$5^{\prime}$ in the $x^{7,8}$ plane, by a quantity proportional
to the derivative of the superpotential: $x^{6}+ix^{7}\propto\mathcal{W}%
^{\prime}(x^{4}+ix^{5})$. The resulting configuration is that of Figure
\ref{cartoonstandard}. Vortices correspond to D$2$-branes stretched between a
color-flavor locked D$4$-brane and the NS$5^{\prime}$-brane.%

\begin{figure}
[ptbh]
\begin{center}
\includegraphics[
height=2.4076in,
width=3.4757in
]%
{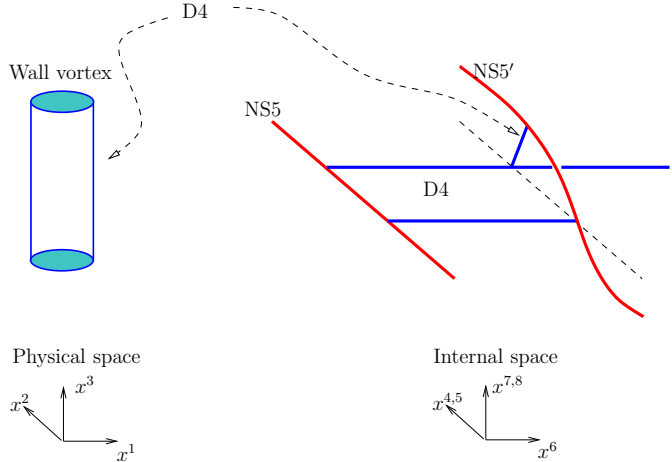}%
\caption{{\protect\footnotesize A lot of coincident D$2$-branes expand out in
a D$4$-brane. In the physical space this corresponds to the wall vortex. }}%
\label{cartoonwallvortex}%
\end{center}
\end{figure}

When a lot of vortices are close to each other we obtain the wall vortex. We
expect that even in this case there is a simple classical description in terms
of D-branes. Our proposal is the following. The D$2$-branes expand out and get
transformed into a D$4$-brane like in Figure \ref{cartoonwallvortex}. This
D$4$-brane is extended in the time direction, the three physical dimensions of
the wall vortex, and the segment between the locked D$4$-brane and the
NS$5^{\prime}$-brane. Inside the wall vortex the locked D$4$-brane is
connected to the NS$5^{\prime}$-brane and so the $U(1)$ gauge is restored.
This explains the presence of the magnetic flux in core of the wall vortex.%

\begin{figure}
[tbh]
\begin{center}
\includegraphics[
height=1.6994in,
width=4.2263in
]%
{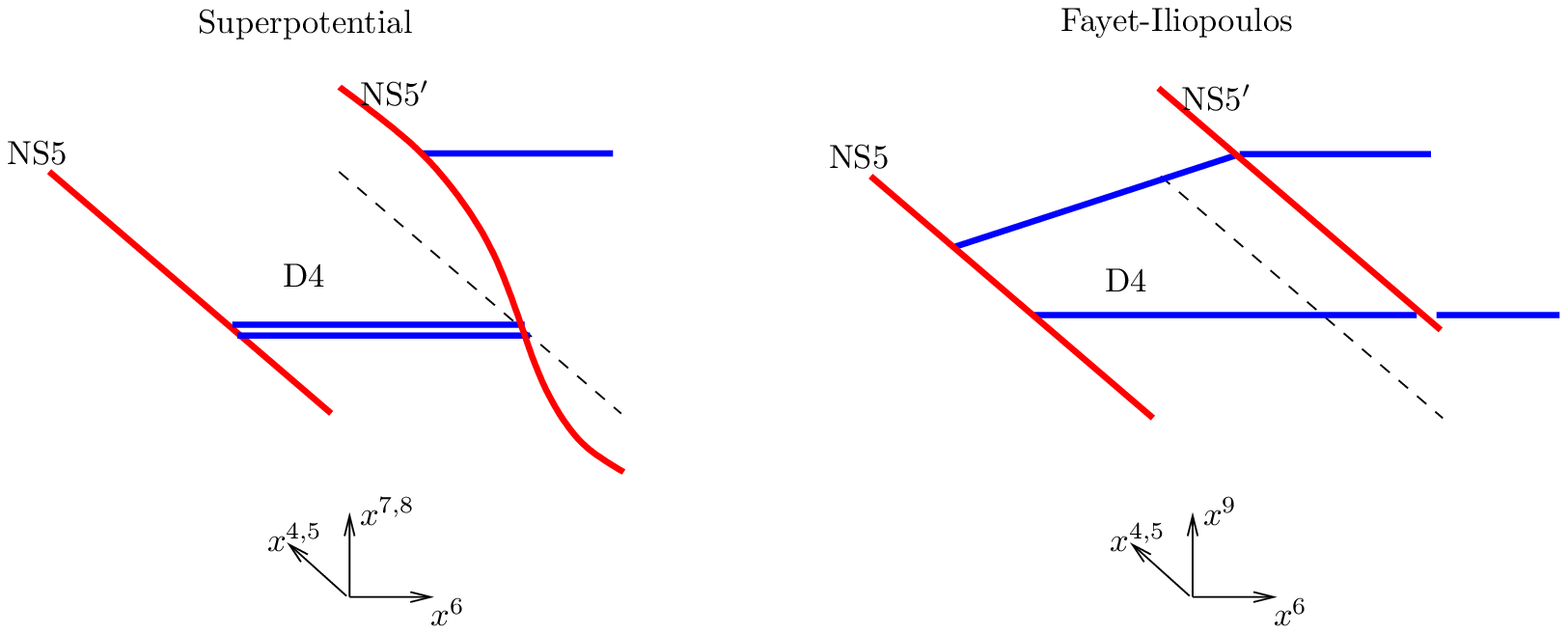}%
\caption{{\protect\footnotesize The locked D$4$-brane of Figure
\ref{cartoonwallvortex} wants to minimize its energy. Two cases must be
distinguished: the breaking by a superpotential when there is at least one
root of $W^{\prime}$ and the breaking by a Fayet-Iliopoulos term.}}%
\label{minimization}%
\end{center}
\end{figure}

Figure \ref{cartoonwallvortex} is not the end of the story. Now that the
D$4$-brane is reconnected to the NS$5^{\prime}$-brane, it tries to minimize
its energy. Two cases must be distinguished. If the $\mathcal{N}=1$ breaking
is obtained by a superpotential, the D$4$-brane splits in two pieces, one
reaches the nearest root of $\mathcal{W}^{\prime}(x^{4}+ix^{5})$, and the
other remains attached to the NS$5^{\prime}$-brane (see first part of Figure
\ref{minimization}). If the $\mathcal{N}=1$ breaking is due to a
Fayet-Ilipoulos term (or equivalently to a linear superpotential), the
D$4$-brane is lifted as in the second part of Figure \ref{minimization}.%

We conclude the section by making a comparison between the string
interpretations of the two bag solitons.

\begin{itemize}
\item When a lot of D$1$-branes ($1$-monopoles) a near to each other they get
transformed into a D$3$-brane (magnetic bag) with a magnetic flux turned on.
The two lacking dimensions are given by the bag surface $\mathcal{S}%
_{\mathrm{m}}$.

\item When a lot of D$2$-branes ($1$-vortices) are close to each other they
become a D$4$-brane (wall vortex) with a magnetic flux turned on. The two
lacking dimensions correspond to the area of the wall vortex $\mathcal{A}%
_{\mathrm{V}}$.
\end{itemize}

These two phenomena are very similar and in fact, if we lift both
configurations to M-theory ($x^{10}$ is compactified on a circle), we discover
that they are identical. A lot of M$2$-branes near to each other get
transformed into a M$5$-brane wrapped on the M-theory circle. On the
M$5$-brane there is a flux $F_{10,i,j}$ turned on where $F$ is the field
strength of the bi-form $A$ that lives on the M$5$-brane. The field
$F_{10,i,j}$ is nothing but the magnetic flux that passes through the bag
surface $\mathcal{S}_{\mathrm{m}}$, in the case of the magnetic bag, and
$\mathcal{A}_{\mathrm{V}}$ in the case of the wall vortex.

\section{Multi-monopoles and Gauge Theories\label{multiandgauge}}

In this section we want to explore the relation between the 't Hooft large $n$
limit of certain $SU(n)$ gauge theories and the large $n$ limit of
multi-monopoles discussed in this paper.

Let us start with the simplest case. It has been discovered in
\cite{Seiberg:1996nz,Chalmers:1996xh} that the Coulomb branch of
$\mathcal{N}=4$ $SU(n)$ Yang-Mills theory in $d=2+1$ dimensions is isomorphic,
as a hyper-Kahler manifold, to the moduli space of $n$ uncentered BPS
monopoles of a $SU(2)$ gauge theory in $d=3+1$ dimensions. A simple
explanation of this has been found by Hanany-Witten \cite{Hanany:1996ie}
considering a brane configuration of $n$\ horizontal\ D$3$-branes suspended
between two vertical NS$5$-branes. Looking at the low energy limit of this
configuration from a \textquotedblleft horizontal\textquotedblright%
\ perspective we recover the $SU(n)$ three dimensional gauge theory while,
from a \textquotedblleft vertical\textquotedblright\ perspective, we recover
the $SU(2)$ gauge theory with $n$ monopoles. The moduli space under
consideration has real dimension $4(n-1)$, i.e. four times the rank of the
$SU(n)$ gauge group.

Now we want to make the large $n$ limit of the three dimensional gauge theory.
According to the 't Hooft prescription, while sending $n$ to infinity we have
also to rescale the gauge coupling $\frac{1}{g_{YM}^{2}}\sim n$ or,
equivalently, keep the dynamical scale $\Lambda$ fixed. But in a theory such
as the one under consideration, we have also to specify the sequence of points
in the moduli space of vacua. The main problem is that the moduli space changes
as $n$ change and so it is not obvious what sequence must be chosen in order to
obtain a definite limit. In the theory under consideration the moduli space is
that of $n$ BPS monopoles so our guess is that we have to use the prescription
adopted in this paper: sending $n$ to infinity while keeping fixed the shape
of the magnetic bag. Of course in order to properly define this limit we have
to understand the large $n$ limit of multi-monopoles and what is discussed in
Subsection \ref{jarvis} are just some speculations along this direction.

But one thing we at least can check: the 't Hooft rescaling correspond exactly
to the rescaling used in Section \ref{moreonthebag} where the radius of the
magnetic bag is kept fixed while making the large $n$ limit. We have to
consider the parameters that enter in the Type IIB brane configuration. We
have the string coupling constant $g_{s}$, the Regge slope $\alpha^{\prime}$
which we set to one for convenience, the distance between the two NS$5$-branes
$v$ and the number of D$3$-branes $n.$ The three dimensional $SU(n)$ gauge
theory knows only about two combination parameters, the size of the gauge
group $n$ and the coupling constant $\frac{1}{g_{YM}^{2}}=\frac{v}{g_{s}}$.
\ The four dimensional $SU(2)$ gauge theory knows about three parameters: the
coupling constant $\frac{1}{e^{2}}=\frac{1}{g_{s}}$, the vev of the Higgs
field $v$ and the number of monopoles $n$. If we want to perform the 't Hooft
large $n$ limit we have to send $\frac{1}{g_{YM}^{2}}\sim n$ to infinity. From
the string theory embedding we can we rescale $v\sim n$ and keep $g_{s}$
constant. In the $SU(2)$ gauge theory this implies that the radius of the
magnetic bag $R_{m}\sim\frac{n}{v}$ remains constant while making the large
$n$ limit.

An interesting generalization of the correspondence between moduli spaces of
gauge theories and magnetic monopoles has been found in \cite{Cherkis:2000cj}.
Here it is shown that the Coulomb branch of $\mathcal{N}=2$ $SU(n)$ Yang-Mills
theory defined on $\mathbf{R}^{3}\mathbb{\times}\mathbf{S}^{1}$ is isomorphic
to the moduli space of $n$ periodic monopoles (to be defined soon). Now we
give some details about this correspondence. The gauge theory is compactified
on a circle of radius $L$. When the radius shrinks to zero we obtain
$\mathcal{N}=4$ $SU(n)$ Yang-Mills theory defined on $\mathbf{R}^{3}$ whose
moduli space has been discussed previously. For general radius the moduli
space is always $4(n-1)$ real dimensional.\footnote{In the decompactification
limit ($R\rightarrow\infty$) the theory becomes a $\mathcal{N}=2$ $SU(n)$
Yang-Mills theory in $d=3+1$. The Coulomb branch of this theory, with all
quantum corrections, has been determined by the Seiberg-Witten solution. In
the decompactification limit there is a discontinuity in the dimension of the
moduli space which is $2(n-1)$ real dimensional.} The periodic monopoles are
solutions of the Bogomol'nyi equations on $\mathbf{R}^{2}\times\mathbf{S}^{1}$
where the radius of the circle is $\widetilde{L}=\frac{1}{L}$. We can think of
them as solutions of the Bogomol'nyi equations in $\mathbf{R}^{3}$ and
periodic in the $x_{3}$ direction with period $2\pi\widetilde{L}$. Periodic
monopoles have been further investigated in \cite{Ward:2005nn} and
\cite{Dunne:2005pr}. The first important feature of periodic monopoles is that
they are not finite in energy. This has to do with the divergence of a charge
in two dimensions at spatial infinity. Another feature, that is related to the
previous one, is that the Higgs field cannot approach a constant value at
$\left\vert s\right\vert \rightarrow\infty$ but it diverges logarithmically.
(This is also related to the logarithmical running of the gauge coupling at
high energy, at high energy the theory is four dimensional.) \ The boundary
condition is thus:%
\begin{equation}
\phi\sim\frac{n}{2\pi\widetilde{L}}\log\left(  \left\vert s\right\vert
\Lambda\right)  +\dots\label{boundary}%
\end{equation}
where $s=x_{1}+ix_{2}$, $\phi$ is the norm of the Higgs field and the ellipses
stand for subleading corrections. For dimensional reasons we
have to introduce a scale $\Lambda$ in Eq.~(\ref{boundary}). From  now on we call $\Lambda$ dynamical scale.
 The coefficient $\frac{n}{2\pi\widetilde{L}%
}$ is determined by the Bogomol'nyi equations and is equal to the linear
charge density of the $n$ periodic monopoles.

We now discuss what we expect to happen to the $n$ periodic multi-monopole in
the $\widetilde{L}\rightarrow0$ limit that correspond to decompactification
limit of the related gauge theory. Dealing with periodic multi-monopoles there
is an important dimensionless quantity to consider $2\pi\widetilde{L}\Lambda$,
i.e. the ratio between the period $2\pi\widetilde{L}$ and the inverse of the
dynamical scale.

 First  we discuss the spherical multi-monopole. If $2\pi\widetilde
{L}\Lambda\gg1$ we have a chain of spherical magnetic bags and the distance
between them is much larger than their radius that is
\begin{equation}
R_{\mathrm{m}}=\frac{n}{v}\simeq\frac{2\pi\widetilde{L}}{\log\left(
2\pi\widetilde{L}\Lambda\right)  }\ll2\pi\widetilde{L}~.
\end{equation}
As $2\pi\widetilde{L}\Lambda$ is decreased the radius of the magnetic bags
will increase until they touch and they will form a unique surface separating
the internal phase from the external one. When $2\pi\widetilde{L}\Lambda\ll1$
it is easy to imagine that the resulting configuration will be that of a
\emph{magnetic tube}. To determine the radius $R_{\mathrm{t}}$ of the tube we cannot
use a minimization energy approach as we have used for the magnetic bag since
the energy is now infinite. But we can solve the problem combining the abelian
Bogomol'nyi equations $\vec{\nabla}\varphi=\vec{\nabla}\phi$ and the boundary
condition (\ref{boundary}) (here we are using the same conventions of Section
\ref{modulispace}, $\varphi$ is the magnetic scalar potential and $\phi$ is
the norm of the Higgs field). The norm of the Higgs field is completely
determined by the following conditions: it must vanish at the radius
$\left\vert s\right\vert =R_{\mathrm{t}}$, it must be an harmonic function and
at infinity must approach (\ref{boundary}). Thus we obtain%
\begin{equation}
\phi=\frac{n}{2\pi\widetilde{L}}\log\left(  \left\vert \frac{s}{R_{\mathrm{t}%
}}\right\vert \right)  ~,\label{tube}%
\end{equation}
and the radius of the tube is the inverse of the dynamical scale
$R_{\mathrm{t}}=\frac{1}{\Lambda}$. The 't Hooft scaling of the coupling constant, that correspond to $v\sim n$, does not
change the parameter $2\pi\widetilde{L}\Lambda$ and neither the size or the shape of the surface.

It is interesting to discuss another configuration, that of axial symmetric
periodic multi-monopoles. We point the ax of the axial symmetric
multi-monopoles in the $x_{1}$ direction. We thus have a chain of magnetic discs
located at $(0,0,k\cdot2\pi\widetilde{R})$ that lie in the $(x_{2},x_{3})$
plane. If $2\pi\widetilde{L}\Lambda\gg1$ we have a chain of magnetic discs and
the distance between them is much larger than their radius
\begin{equation}
R_{\mathrm{d}}=\frac{\pi n}{2v}\simeq\frac{4\pi\widetilde{L}}{\pi\log\left(
2\pi\widetilde{L}\Lambda\right)  }\ll2\pi\widetilde{L}~.
\end{equation}
When $2\pi\widetilde{L}\Lambda\ll1$ the configuration becomes a \emph{magnetic strip}
described by $x_{1}=0$, $-R_{\mathrm{s}}\leq x_{2}\leq R_{\mathrm{s}}$ and
$-\infty\leq x_{3}\leq\infty$. The field $\phi$ must be an harmonic function
that vanishes on the strip and has the boundary conditions (\ref{boundary}).
The solution is
\begin{equation}
\phi=\frac{n}{2\pi\widetilde{L}}\operatorname{Re}\cosh^{-1}\left(  \frac
{s}{R_{\mathrm{s}}}\right)  \label{strip}%
\end{equation}
and the thickness of the strip is twice the radius of the tube $R_{\mathrm{s}%
}=\frac{2}{\Lambda}$. The same result (\ref{strip}) has been obtained in
\cite{Ward:2005nn}\footnote{Here $n=1$ and the distance $\widetilde{R}$ is
sent to zero. The result is still a magnetic strip because in a chain of
periodic single monopoles they have all the same $U(1)$ phase factor. When two
monopoles with the same $U(1)$ phase collide they create an axial symmetric
multi-monopole.}.

\section{Multi-monopoles and Cosmology \label{cosmology}}

If we want to make a phenomenological discussion about magnetic monopole, we
inevitably have to confront with cosmology. In fact magnetic monopoles, if
they exist, arise as stable solitons of the grand unification symmetry
breaking. If their masses are of order $10^{16}\,\text{GeV}$, the only way
they can be produced is in the cosmological contest, when the temperature of
the universe was of order of the\ GUT scale. In this section we want to
address the following question:

\begin{itemize}
\item Is it possible to have multi-monopole formation in the cosmological context?
\end{itemize}

The answer is yes if we can play with some parameters of the GUT Higgs
potential. The only two parameters that enter in the game are $V(0)$, the
value of the Higgs potential in zero, and $V^{\prime\prime}(v)$ that
corresponds to the mass of the neutral Higgs boson. We will also see that our
mechanism for the production of multi-monopoles brings also another
consequence: a reduction of the total number of monopoles plus anti-monopoles.
We thus can address another question:

\begin{itemize}
\item Is it possible to solve the monopole cosmological problem?
\end{itemize}

The answer is again yes, but now we have to make an extreme choice of the
parameter $V^{\prime\prime}(v)$.

In the Subsection \ref{monopolesandcosmology} we briefly review the ordinary
theory of the monopole production during the GUT phase transition. In
Subsection \ref{possiblemechanism} we provide a possible mechanism for the
formation of multi-monopoles and finally in Subsection \ref{monopoleproblem}
we discuss the possible solution of the cosmological monopole problem.

\subsection{Monopoles and Cosmology \label{monopolesandcosmology}}

\begin{figure}
[tbh]
\begin{center}
\includegraphics[
height=2.2174in,
width=3.0407in
]%
{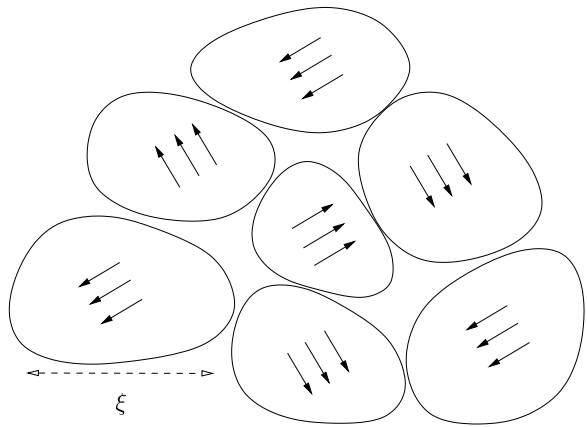}%
\caption{{\protect\footnotesize The Kibble mechanism. In any cosmological
phase transition the order parameter is correlated in domains of finite length
$\xi$. The length $\xi$ must be finite since it is bounded from below by the
horizon length $d_{H}$.}}%
\label{kibble}%
\end{center}
\end{figure}

\begin{figure}
[tbh]
\begin{center}
\includegraphics[
height=2.4422in,
width=3.1332in
]%
{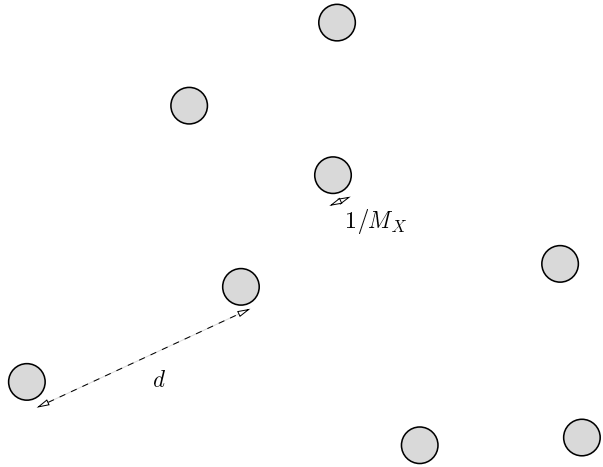}%
\caption{{\protect\footnotesize At the intersection of the various domains in
Figure \ref{kibble}, there is a probability $p$ of finding a topological
defect. The outcome of the phase transition is thus a homogeneous distribution
of monopoles and anti-monopoles with size $1/M_{X}$ and mean distance $d$,
where $d^{-3}\sim p\xi^{-3}$. }}%
\label{monopoli}%
\end{center}
\end{figure}

Now we briefly recall the cosmological monopole production (see
\cite{Preskill:1984gd} for a review). When the universe cooled below the
critical temperature of the GUT phase transition $T_{c}$, the scalar field
$\phi$ condensed in various domains of length $\xi$ (see Figure \ref{kibble}).%
The finiteness of the length $\xi$ is the crucial point for the existence of
topological defects \cite{Kibble:1976sj}. Even if the correlation length
becomes infinite at the critical temperature, the length $\xi$ is always
bounded from below by the horizon scale $d_{H}$. At the intersection of the
various domains there is a probability $p$ of finding a topological defect and
$p\sim1/10$ in the case of the monopole. This implies that we can neglect the
production of multi-monopoles at this stage. The outcome of the phase
transition is presented in Figure \ref{monopoli}. We have a distribution of
single monopoles of density $d^{-3}\sim p\xi^{-3}$.%

Let's take a look at the various order of magnitudes in the problem. The mass
of the heavy GUT bosons is $M_{X}\sim10^{15}\,\text{GeV}$ and is almost the
same as the critical temperature $T_{c}$. At this temperature the horizon
scale is ${d_{H}}\sim(10^{12}\,\text{GeV})^{-1}$. If we take the extreme case
in which the Kibble bound is saturated, we thus have a distribution of
monopoles and anti-monopoles of typical distance $d\sim(10^{12}\,\text{GeV}%
)^{-1}$ and radius ${R_{\mathrm{m}}}\sim{M_{X}}^{-1}\sim(10^{15}%
\,\text{GeV})^{-1}$. Generally the mass of the GUT Higgs boson is considered
of the same order of the $X$ boson. Thus our configuration satisfy the
following conditions:%
\begin{equation}
R_{\mathrm{m}}\sim\frac{1}{M_{X}}\sim\frac{1}{M_{H}}\ll d\ . \label{ineq}%
\end{equation}
In this regime we can treat the system as a neutral plasma of monopoles and
anti-monopoles whose interaction is only due to the magnetic field. At this
stage the only physical effect that can happen is the annihilation between
monopoles and anti-monopoles. Since the system is neutral, there is only a
small drift force that cause the attraction between them. The calculations in
\cite{Zeldovich:1978wj, Preskill:1979zi} show that this process, when we take
into account also the expansion of the universe, is essentially negligible.
The predicted density of monopoles and anti-monopoles is many order of
magnitudes bigger than the upper bound posed by the observations. This is the
so called cosmological monopole problem, an enormous discrepancy between the
prediction of the GUT models and the observational bounds.

A lot of possible solutions have been proposed to this problem. One is in the
context of inflation \cite{Guth:1980zm,Linde:1981mu}. If the GUT phase
transition is before or during inflation, the density of monopoles and
anti-monopoles can be enormously diluted by the exponential expansion of the
universe. Another possible solution is that the universe undergoes a
intermediate phase transition where the electromagnetic $U(1)$ is in the Higgs
phase \cite{Langacker:1980kd}. In this phase monopoles and anti-monopoles are
confined by flux tubes and the annihilation process is enhanced. More recent
speculations on the subject are considered in Refs. \cite{Dvali:1995cj}.

\subsection{A mechanism for the formation of multi-monopoles
\label{possiblemechanism}}

\begin{figure}
[tbh]
\begin{center}
\includegraphics[
height=1.8075in,
width=4.3163in
]%
{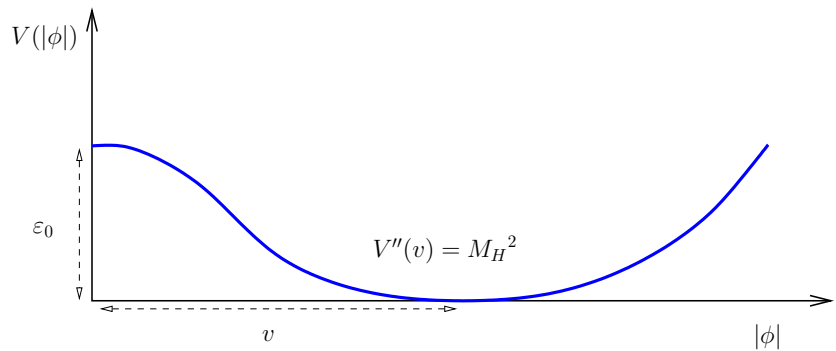}%
\caption{{\protect\footnotesize A potential that can lead to the production of
multi-monopoles.}}%
\label{potentialadhoc}%
\end{center}
\end{figure}

As we previously mentioned, the production of multi-monopoles can be
considered negligible in the usual scenario. If we want to explore the
possibility of the production of multi-monopoles, we need to change the
potential for the GUT symmetry breaking. In particular there are two
parameters that play a fundamental role: the zero energy density
$V(0)=\varepsilon_{0}$ and the Higgs boson mass $M_{H}=V^{\prime\prime}(v)$
(see Figure \ref{potentialadhoc}).

First of all we evaluate the spectrum of multi-monopoles using the results of
Section \ref{bag}. We just have to plot, like in Figure \ref{spettro}, the BPS
mass and the MIT bag mass%
\begin{equation}
M_{\mathrm{BPS}}(n)=\frac{4\pi v}{e}\,n\ ,\qquad M_{\mathrm{MIT}}%
(n)=\frac{2^{1/4}7\pi}{3}\,\frac{{\varepsilon_{0}}^{1/4}}{e^{3/2}}\,n^{2/3}\ ,
\end{equation}
and they intersect at%
\begin{equation}
n^{\ast}\sim2e\frac{v^{2}}{\sqrt{\varepsilon_{0}}}\ . \label{regionmarginal}%
\end{equation}
Far away from the transition between the two regimes, we can approximate the
mass as $M_{\mathrm{m}}=\max{(M_{\mathrm{BPS}},M_{\mathrm{MIT}})}$. Taking
$\varepsilon_{0}$ sufficiently smaller than $v^{4}$, there is a long BPS
region between $1<n<n^{\ast}$, then a small transition between the two regimes
and finally the MIT bag regime $n>n^{\ast}$. Up to the value $n^{\ast}$ we can
have marginally stable multi-monopoles, above that number the multi-monopoles
will be unstable to decay into multi-monopoles of magnetic charge lower than
$n^{\ast}$. From (\ref{regionmarginal}) we see that we the smaller
$\varepsilon_{0}$ is, the more the region of marginal stability is wide.

\begin{figure}
[tbh]
\begin{center}
\includegraphics[
height=3.5526in,
width=3.7455in
]%
{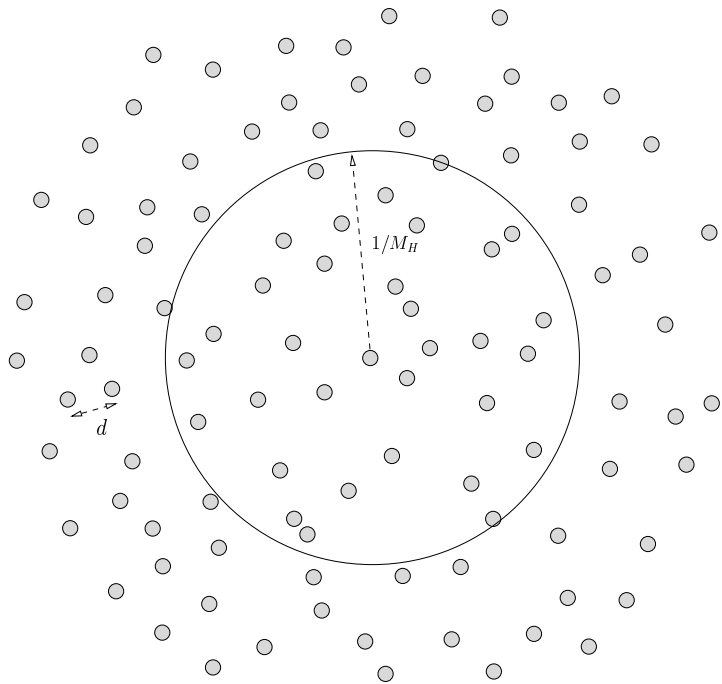}%
\caption{{\protect\footnotesize Inside the sphere of radius $1/M_{H}$ the correct
approximation is that of a plasma of BPS monopoles.}}%
\label{statistica}%
\end{center}
\end{figure}

If we want to change the evolution of monopoles after the phase transition, we
need to change something in the inequalities (\ref{ineq}). What we want is a
regime in which the following inequalities are satisfied:%
\begin{equation}
R_{\mathrm{m}}\sim\frac{1}{M_{X}}\ll d\ll\frac{1}{M_{H}}\ , \label{ineqnew}%
\end{equation}
and the situation is described in Figure \ref{statistica}.%
Consider a sphere of radius $1/M_{H}$ that is in the regime (\ref{ineqnew}),
the radius is much larger than the mean distance $d$. Inside this sphere we
cannot approximate the monopoles as a plasma of particles interacting only
with the magnetic field. The correct approximation is that of a plasma of BPS
monopoles, where the force due to the exchange of the Higgs boson gives a
fundamental contribution to the dynamics. In particular the force between two
BPS monopoles is zero while the force between a monopole and an anti-monopole
is doubly attractive \cite{Manton:1977er}. So the physics inside a sphere of
radius $1/M_{X}$ is very similar that of a system of particles with
gravitational interaction, we have only $1/r^{2}$ attractive forces and no
repulsive forces. It is known from the theory of structure formation, that the
expansion of the universe cannot stop the gravitational collapse but only
change its dependence on time from exponential to polynomial.

In the theory of gravitational instability \cite{Coles:1995bd} there is a
particular quantity to take care of: the Jeans length $\lambda_{J}$. If the
scale of the density fluctuation is greater than $\lambda_{J}$ we can have a
collapse, otherwise the fluctuation will continue to oscillate without
growing. The collapse of the sphere in Figure \ref{statistica} can happen only
if $1/M_{H}$ is greater than the Jeans length. A crude estimate of
$\lambda_{J}$ can be obtained as follows. We take a sphere of radius $\lambda$
and we consider a particle at the edge of the sphere. The potential energy of
the particle is $\sim N/\lambda$ where $N\sim(\lambda/d)^{3}$ is the number of
particles in the sphere. The Jeans length is the one at which the potential
energy becomes comparable to the kinetic energy. Above this scale the
potential energy dominates over the dissipation and the clustering can happen.
Taking $d\sim(10^{12}\,\text{GeV})^{-1}$ and the kinetic energy $10^{15}%
\,\text{GeV}$ we obtain ${\lambda_{J}}\sim(10^{9}\,\text{GeV})^{-1}$. We thus
have the constraint $M_{H}<10^{9}\,\text{GeV}$ if we want the collapse to take place.

At this point we need to say something about the statistical behavior of
monopole and anti-monopoles. We call $n_{+}$ and $n_{-}$ the number of
monopoles and antimonopoles inside a sphere of radius $1/M_{H}$. The number of
particles is%
\begin{equation}
n=n_{+}+n_{-}\sim\frac{1}{(M_{H}d)^{3}}\ .
\end{equation}
The total charge is $\delta n=n_{+}-n_{-}$. If we consider all the possible
spheres of radius $1/M_{H}$, $\delta n$ is a stochastic variable with zero
mean $\langle\delta n\rangle=0$. The physical quantity we are interested on is
the variance $\sqrt{\langle\delta n\rangle}$ that, from now on, we denote for
simplicity $\delta n$. Suppose for a moment that the $n$ particles inside the
sphere are independent stochastic variables. Every particle can assume the
value $+1$ (monopole) or $-1$ (anti-monopole) with equal probability. This
would give a variance $\delta n\sim\sqrt{n}$. This naive expectation is wrong
since the particles are strongly correlated. The total magnetic charge can in
fact be expressed as an integral over the surface of the sphere
\cite{Arafune:1974uy}%
\begin{equation}
\delta n=\frac{1}{8\pi}\int ds^{ij}\frac{\epsilon_{abc}\phi^{a}\partial
_{i}\phi^{b}\partial_{j}\phi^{c}}{|\phi|^{3}}\ .
\end{equation}
This implies that the magnetic charge fluctuation is a surface effect and not
a volume effect. The variance is thus the square root of the surface area
\cite{Vilenkin}%
\begin{equation}
\delta n\sim\frac{1}{M_{H}d}\ .
\end{equation}

The process we have just described has brought another (in principle
unrequested) consequence: a reduction of the total number of monopoles and
anti-monopoles. Before the collapse we had a number $1/(M_{H}d)^{3}$ of
monopoles and anti-monopoles. At the end of the collapse we have a big
multi-monopole of charge $1/(M_{H}d)$. The total number of particles has been
reduced by a factor $1/(M_{H}d)^{2}$.

\subsection{The cosmological monopole problem \label{monopoleproblem}}

It is interesting to ask if this process could solve the cosmological monopole
problem. In principle it could work, we just have to take the mass of the GUT
Higgs small enough. If we take $d\sim(10^{12}\,\text{GeV})^{-1}$ and
$M_{H}\sim10^{3}\,\text{GeV}$ we have a suppression of $10^{18}$, that is
enough to solve the monopole problem. But now the monopole problem is
translated into a hierarchy problem, the mass of the GUT Higgs boson is
considerably lighter than the GUT scale. This is not so bad if this hierarchy
has same relation with the electroweak hierarchy. It is believed that at
$10^{3}\,\text{GeV}$ some new physics will be discovered. This new physics
(technicolor, supersymmetry or something that is still unknown) should explain
why the electroweak scale is so small compared to the GUT scale.

To summarize we have the following scenario. We have a GUT scale at
$10^{15}\,\text{GeV}$ and some \textquotedblleft protection\textquotedblright%
\ at $1\,\text{TeV}$ that solves the electroweak hierarchy problem. Now
suppose that the GUT Higgs boson is essentially massless at the GUT scale and
acquires mass only at $1\,\text{TeV}$. After the GUT phase transition a
certain distribution of monopoles and anti-monopoles is created. The
subsequent evolution of these particles is usually described by a neutral
plasma of charged particles. This approximation does not work in our scenario.
The GUT Higgs boson gives an essential contribution and, inside a sphere of
radius $1/M_{H}$, we can approximate our system as a plasma of almost-BPS
monopoles. The physics of this system is very similar to that of a plasma of
gravitational interacting particles and the collapse, even if it is slowed due
to the expansion of the universe, is unavoidable. Our model predicts a
reduction of the number of monopoles and anti-monopoles of order
$1/(M_{H}d)^{2}\sim10^{-18}$. This number is big enough to solve the monopole
problem and small enough to leave us the chance to observe magnetic monopoles.

\section*{Acknowledgements}

I thank J.~Evslin, S.~B.~Gudnason, K.~Konishi, F.~Sannino and D.~Tong for
comments and discussions. This work is supported by the Marie Curie Excellence
Grant under contract MEXT-CT-2004-013510.

\end{document}